\let\latexaddtocontents\addtocontents
\documentclass[submitting, floatfix]{nst}
\let\addtocontents\latexaddtocontents
\usepackage{subfigure,dcolumn}
\usepackage[T2A,T1]{fontenc}
\usepackage[english]{babel}
\usepackage{color}
\usepackage{listings}

\usepackage{natbib}
\let\footnote\relax

\let\textcite\relax

\let\citeauthor\relax
\let\citeyear\relax
\expandafter\let\csname
ver@natbib.sty\endcsname\relax

\usepackage[
    backend=biber,
    bibencoding=utf8,
    style=numeric-comp,
    giveninits=true,
    autolang=other, 
    sorting=none,
    doi=true, isbn=false, eprint=false, url=false, date=year, 
    maxcitenames=3,
    mincitenames=3,
    maxbibnames=3,
    minbibnames=3,
]{biblatex}
\begin{document}
\renewcommand{\bibliography}[1]{}
\title{Method for detector description conversion from DD4hep to Filmbox}
\thanks{This work was supported by the National Natural Science Foundation of China (Grant Nos. 12175321, 11975021, 11675275, and U1932101), National Key Research and Development Program of China (Nos. 2023YFA1606000 and 2020YFA0406400), State Key Laboratory of Nuclear Physics and Technology, Peking University (Nos. NPT2020KFY04 and NPT2020KFY05), Strategic Priority Research Program of the Chinese Academy of Sciences (No. XDA10010900), National College Students Science and Technology Innovation Project, and Undergraduate Base Scientific Research Project of Sun Yat-sen University.}

\author{Zhao-Yang Yuan}
\author{Tian-Zi Song}
\author{Yu-Jie Zeng}
\author{Kai-Xuan Huang}
\affiliation{School of Physics, Sun Yat-sen University, Guangzhou 510275, China}
\author{Yu-Mei Zhang}
\email[Yu-Mei Zhang, ]{zhangym26@mail.sysu.edu.cn}
\affiliation{Sino-French Institute of Nuclear Engineering and Technology, Sun Yat-sen University, Zhuhai 519082, China}
\author{Zheng-Yun You}
\email[Zheng-Yun You, ]{youzhy5@mail.sysu.edu.cn}
\affiliation{School of Physics, Sun Yat-sen University, Guangzhou 510275, China}

\begin{abstract}

DD4hep serves as a generic detector description toolkit recommended for offline software development in next-generation high-energy physics~(HEP) experiments. 
Conversely, Filmbox~(FBX) stands out as a widely used 3D modeling file format within the 3D software industry.
In this paper, we introduce a novel method that can automatically convert complex HEP detector geometries from DD4hep description into 3D models in the FBX format. 
The feasibility of this method was demonstrated by its application to the DD4hep description of the Compact Linear Collider detector and several sub-detectors of the super Tau-Charm facility and circular electron-positron collider experiments. 
The automatic DD4hep--FBX detector conversion interface provides convenience for further development of applications, such as detector design, simulation, visualization, data monitoring, and outreach, in HEP experiments.

\end{abstract}
 

\maketitle

\nolinenumbers
\section{Introduction}

In the realm of high-energy physics~(HEP) software, with increasing detector complexity and data amounts in next-generation experiments, challenges arise from detector description, high-performance computing applications, data processing, and analysis~\cite{Roadmap}. 
Detector description is an essential part of the HEP software across different stages of the lifetime of an experiment. It provides vital detector information for various applications, such as detector design, optimization, simulation, reconstruction, event display, data monitoring, and physical analysis.

The detector description toolkit for HEP (DD4hep)~\cite{dd4hep_package}, a generic detector description toolkit, draws from the experience of the detector description system development for the large hadron collider (LHC) experiments~\cite{Ponce:2003pr} and the linear collider community~\cite{LinearColliderILDConceptGroup-:2010nqx}.  
DD4hep provides consistent detector descriptions from a single source with minimal information to obtain the desired results for simulation, reconstruction, and analysis applications. 
This has been adopted in conceptual design studies on future high-energy colliders.

FBX~\cite{FBX} is a well-known 3D asset-exchange format that facilitates high-fidelity data exchange.
It is one of the most commonly used geometric modeling formats across various industries, supporting seamless sharing and utilization across various digital modeling and content creation programs.
FBX finds extensive application across various domains, including visualization, video and multimedia development, game production, and enhancing functionalities within the realm of virtual reality~(VR)~\cite{VR} and augmented reality~(AR)~\cite{AR}. 
Furthermore, the FBX file format, which uses surface modeling, offers an innovative approach for simulating detector models. 
However, the direct conversion of the detector description from DD4hep to FBX format remains available, hindering the utilization of 3D software into HEP software development within the industry.

In this study, we developed a method to automate the conversion of DD4hep detector description into the FBX format, thereby bridging the next-generation HEP detector geometry and visualization with industrial modeling. For HEP detectors that have been described using DD4hep in offline software, such as the Compact Linear Collider~(CLIC), Future Circular Collider~(FCC), circular electron-positron collider~(CEPC), and super Tau-Charm facility~(STCF) experiments, the interface could directly convert them into FBX formats. This advancements enables the further development and expansion of applications using industrial 3D software.

The remainder of this paper is organized as follows:
In Section~\ref{sec:detector_description}, we introduce detector description with DD4hep in HEP offline software, FBX 3D modeling, and their respective characteristics. 
Section~\ref{sec:methodologies}, presents the conversion method from DD4hep to the FBX format.
In Section~\ref{sec:application}, we discuss its applications in the CLIC and other experiments, as well as its performance, and potential for further developments.
Finally, Section~\ref{sec:summary}provides a summary of the results.   

\section{Detector description}
\label{sec:detector_description}

Currently, different types of detector geometry formats and libraries are used in HEP software toolkits, such as Geant4~\cite{Geant4}, ROOT~\cite{ROOT}, and the geometry description markup language~(GDML)~\cite{GDML}.
Geant4 is a commonly used toolkit to simulate particle interactions with detectors, including the ATLAS~\cite{atlas} and CMS~\cite{CMS} detectors at the LHC~\cite{LHC} and Jiangmen Underground Neutrino Observatory~(JUNO)~\cite{JUNO}. 
ROOT has been used for detector description~\cite{Corliss:2022fat} in the sPHENIX~\cite{Aidala:2012nz} experiment at the relativistic heavy ion collider and in the design of the detector at the electron-ion collider in China~(EicC)~\cite{Anderle:2021wcy}. 
The geometry of Geant4 or ROOT can be applied in the GDML format, which can also be implemented into the offline software of an experiment, as realized in the BESIII~\cite{BESIII_geo} experiment. 
DD4hep is a generic detector description technique driven by the HEP community, for applications in next-generation HEP experiments, such as CLIC~\cite{clic}, FCC~\cite{FCC}, ILC~\cite{ILC}, CEPC~\cite{CEPC}, STCF~\cite{STCF}, and Muon Colliders~\cite{MuonCollider1,MuonCollider2}.
In addition, the Large Hadron Collider beauty~(LHCb) experiment continues to develop from the geometry to the DD4hep toolkit~\cite{lhcb_indd4hep}.

\subsection{DD4hep}

Future collider experiments require advanced software and computing techniques to optimize detector performance and maximize the physics capabilities. 
The design of detector hardware cannot be separated from the detector description in the offline software, which combines simulation and reconstruction to evaluate detector performance.
A common detector description solution benefits each experiment and save resources for software development.
In the HEP software foundation (HSF) community White Paper~\cite{HSF} and the roadmap for HEP software and computing R\&D for the 2020s~\cite{Roadmap}, DD4hep was recommended for detector description in offline software for the next-generation HEP experiments. 

\begin{figure}
    \centering
    \includegraphics[width=0.9\linewidth]{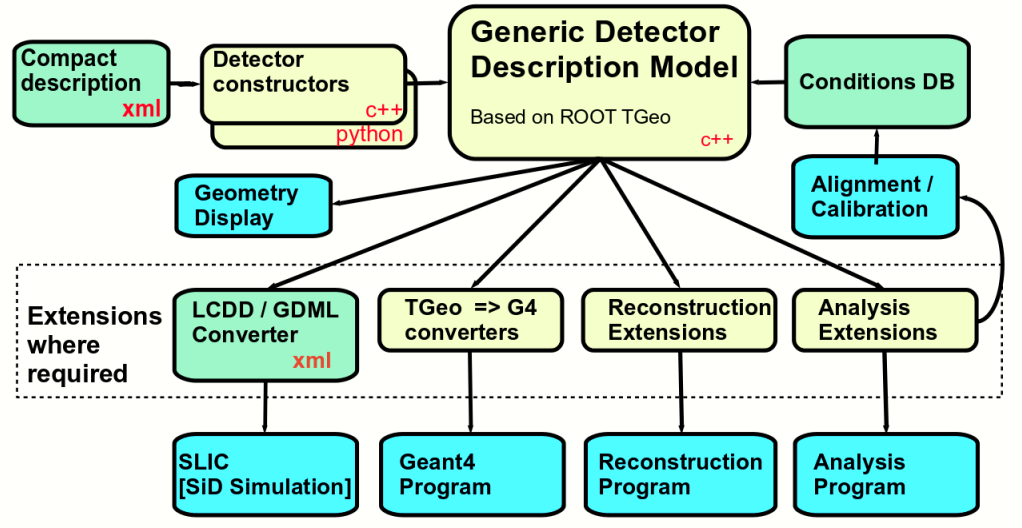}
    \caption{Components of DD4hep detector geometry toolkit~\cite{dd4hep_paper}.}
    \label{fig:dd4hep_toolkit}
\end{figure}

DD4hep provides a comprehensive detector description for the quantification of detectors and the necessary information for the interpretation of geometric data from experiments. 
The structure of DD4hep is shown in Fig. ~\ref{fig:dd4hep_toolkit}.
DD4hep uses a pre-existing and widely used software combined with a consistent generic detector description. 
Its main components include the ROOT geometry package for constructing and visualizing geometry and the Geant4 simulation toolkit, which can interface via DD4hep to perform detector simulations for complex detectors.
DD4hep is designed to operate independently of any Monte Carlo simulation engine. 
This method has the following advantages:
\begin{itemize}
\item \emph{Complete detector description}. It provides comprehensive information on detector geometry, materials, structures, visualization attributes, detector readout information, alignment, calibration, and environmental parameters.
\item \emph{Coverage of the full experiment life cycle}. It supports all stages including initial detector concept development, detector optimization, construction, and operation. Additionally, it enables an easy transition from one stage to another with detector version control.
\item \emph{Single source of information}. It provides a consistent detector description for all applications including simulation, reconstruction, event display, data monitoring, and data analysis.
\item \emph{Ease of Use}. It provides a simple and intuitive interface, with minimal external dependencies.
\end{itemize}

\begin{figure}
    \centering
    \includegraphics[width=0.9\linewidth]{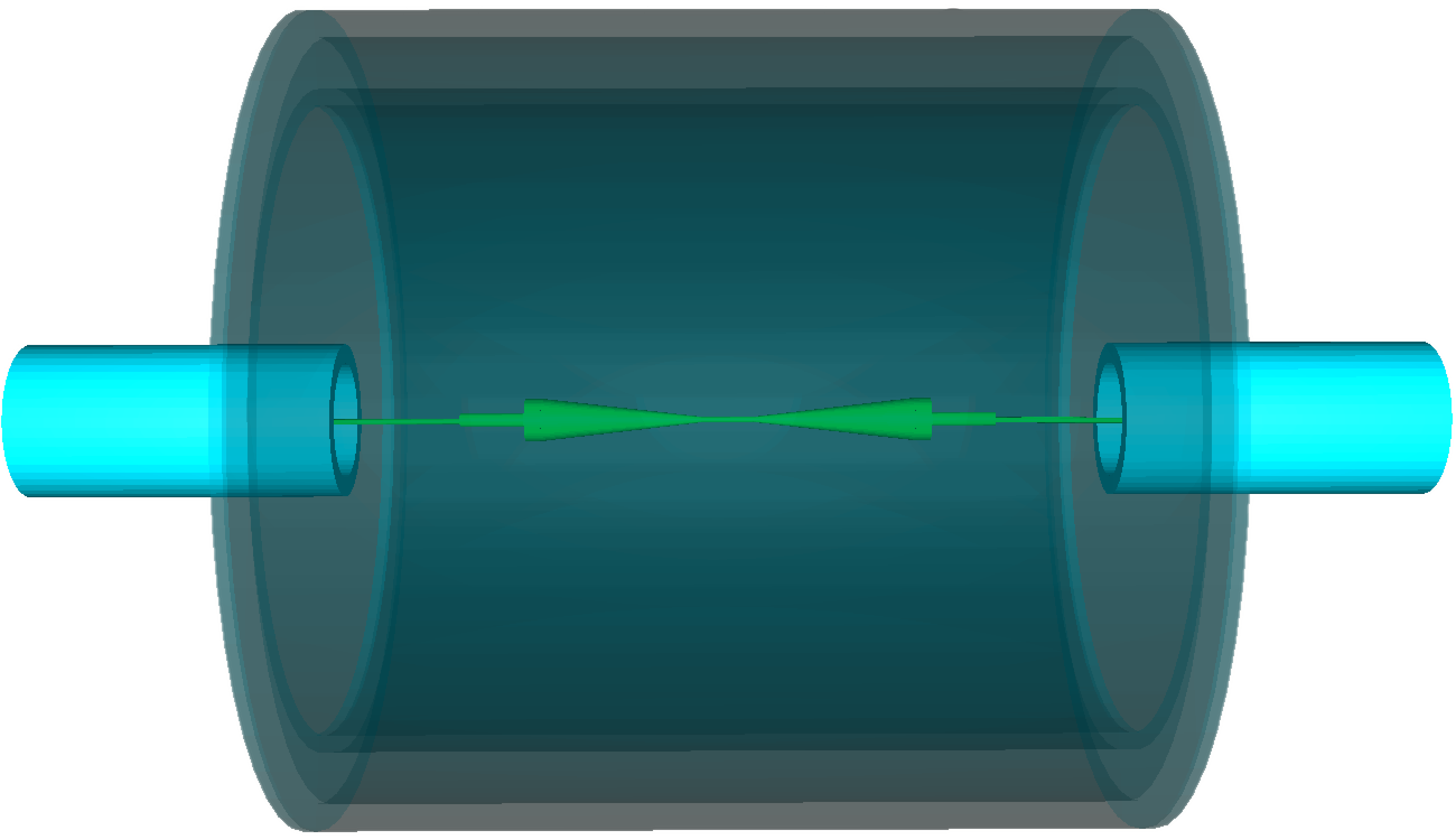}
    \caption{Simple toy detector described with DD4hep and displayed in ROOT.}
    \label{fig:example_dd4hep}
\end{figure}

Fig.~\ref{fig:example_dd4hep} shows a simplified example of the detector geometry description using DD4hep. The detector is visualized using the ROOT OpenGL display engine.

\subsection{FBX format}

\begin{figure}
    \centering
    \includegraphics[width=0.5\linewidth]{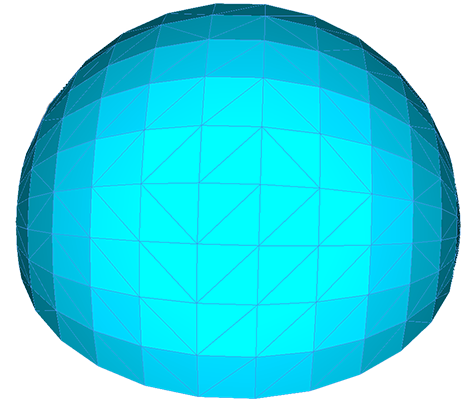}
    \caption{Half sphere described by the mesh in FBX.}
    \label{fig:fbx_mesh}
\end{figure}

FBX, or Filmbox, originally developed by Kaydara for MotionBuilder and acquired by Autodesk in 2006~\cite{maestri2000filmbox,autodesk}, is a 3D-mesh file format used to exchange 3D models and animation data.
It shows potential for industrial applications. 
As shown in Fig.~\ref{fig:fbx_mesh}, FBX describes the geometric structure of objects using a mesh. 
FBX files are compatible with various scenarios and software platforms of unreal Engine, including 3ds Max~\cite{3dmax}, Maya~\cite{maya}, Blender~\cite{Blender}, Unity~\cite{Unity}, and Unreal Engine~\cite{UnReal}. 
FBX can store models, textures, lighting, animations, and special effects. It is often used in the video game industry for character assets and environments.

FBX has wide applications in 3D modeling and animation, extending to areas such as VR and AR. FBX files contain information related to geometries, materials, animations, skeletons, and other data, rendering them potential file formats. 
In Unity, FBX files can be directly imported for building game scenes, character models, and animations. They also supports detection and instance visualization, enabling developers to effectively understand and debug models and animations.

\subsection{Surface description}
\label{sec:poly_surface}

The evolution of HEP depends on highly statistical physical instances. With continuous progress in science and technology, the demand for considerable data and high computing power in experimental physics research has increased. As an important part of experimental calculations, detector simulation requires considerable computational power to simulate the interaction and propagation of particles in a detector.

A new technology is gradually being applied to detector simulation. In other words, graphics processing units~(GPUs) are used rather than the traditional central processing unit~(CPU) to achieve certain simulation functions. Compared with CPUs, GPUs have high parallel computing power and can process considerable amounts of data simultaneously, thus accelerating the simulation process. The application of this technique is expected to improve simulation performance and reduce calculation time in future experiments.

Furthermore, the structural description of a detector affects simulation performance. Traditional detector descriptions, including Geant4, ROOT, and GDML, rely on constructive solid geometry~(CSG), which has limitations in GPU simulations. In contrast, a detector structure employing a polygonal surface description is suitable for GPU simulation because it can efficiently use the parallel computing power of the GPU~\cite{Tognini:2022nmd}.

Therefore, one of the development directions for improving the simulation performance of detectors in the future is to further explore the application of GPU in simulations while optimizing the geometric description of detectors to adapt to the characteristics of GPU parallel computing. This approach will accelerate the processing and analysis of experimental data and advance the development of physics.

\section{Methodologies}
\label{sec:methodologies}

Industrial 3D software, such as Unity~\cite{Unity} and unreal engines ~\cite{UnReal}, show great potential for developing detector and event visualization tools in HEP experiments. 
They have been applied to event displays, data analyses, and outreach software development in several HEP experiments, including ATLAS~\cite{CAMELIA,Bianchi:2019can}, BELLE II~\cite{BelleIIVR}, BESIII~\cite{Huang2022}, and JUNO~\cite{ELAINA, juno_reco1, juno_reco2, juno_reco3}.

However, HEP detectors are usually complex with millions of units, making it challenging to construct a 3D detector model using industrial 3D software.
In the offline software of HEP experiments, the detector description may already exist in certain formats used for detector construction in simulation and reconstruction. As DD4hep has been selected for detector description in next-generation HEP experiments, an interface for converting detector geometry data from DD4hep to 3D modeling in Unity or unreal would highly benefit the HEP software community by leveraging potential 3D software.
Therefore, we propose a novel method for automatically converting the DD4hep detector description into the well-known FBX format, which can be directly imported into 3D software to construct the detector model.

\subsection{Automatic detector geometry conversion}

The automatic conversion of detector description originates from the need to maintain geometric consistency across different applications in HEP offline software.
This can be realized by developing various automatic geometry conversion interfaces between different software packages while maintaining a single source of detector data. 

\begin{figure}
    \centering
    \includegraphics[width=0.9\linewidth]{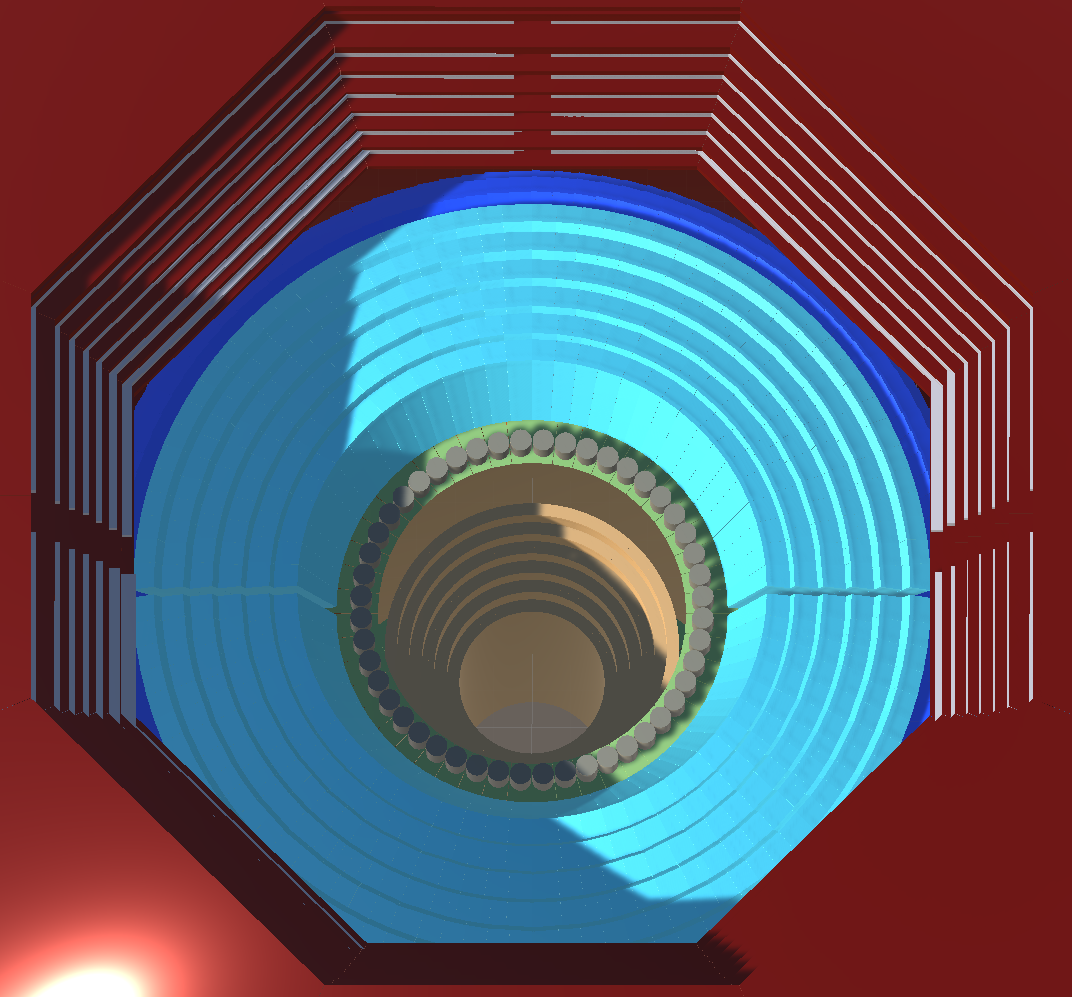}
    \caption{Display of the BESIII detector in Unity with detector geometry data converted from GDML.}
    \label{fig:BesVis3D}
\end{figure}

In the BESIII experiment, the detector description with GDML was used to construct a detector in Geant4 for simulation through the GDML--Geant4 interface, and in ROOT for reconstruction and event display with the GDML--ROOT interface~\cite{BESIII_geo,ROOT_Geant4}.
In addition, a method for detector conversion from GDML to FBX using FreeCAD was proposed to construct a BESIII detector in Unity~\cite{Huang2022} for further event-display development.
Fig.~\ref{fig:BesVis3D} shows the BESIII detector in Unity with the detector geometry converted from GDML, which has excellent visualization effects.
In addition, a tool for directly converting geometry from Geant4 to FBX or VRML format has been developed in the Belle II experiment for detector navigation, physical analysis, and VR applications~\cite{BelleIIVR}.
These interfaces ensure consistency in the detector geometry across simulation, reconstruction, event display, and data analysis. 
Moreover, they save considerable human effort for software developers by providing automatic detector conversion, which enriches the 3D modeling software interface and facilitates the development of extended applications.

As a new-generation detector description technique, DD4hep has a similar architecture to that of automatic detector geometry conversion.
DD4hep was developed using the ROOT geometry package and has an interface with Geant4 detector construction.
In principle, DD4hep can be converted to FBX using existing conversion interfaces, such as the chain of DD4hep--ROOT--GDML--FreeCAD--FBX or DD4hep--Geant4--FBX. 

However, these detector description systems have inherent differences, which can lead to challenges after multiple steps of conversion in a long chain.
Furthermore, these conversions require the construction of a detector first in ROOT or Geant4 during program running, which significantly depends on the Geant4/ROOT geometry construction packages and can cause problems because of different software versions.
If a novel method for direct DD4hep--FBX conversion can be realized, this universal technique would benefit all current and future HEP experiments, enabling the development of applications for detector visualization, event display, and outreach.

\subsection{Detector data conversion from DD4hep to FBX}

\begin{figure}[!htb]
	\includegraphics[width=0.98\hsize]{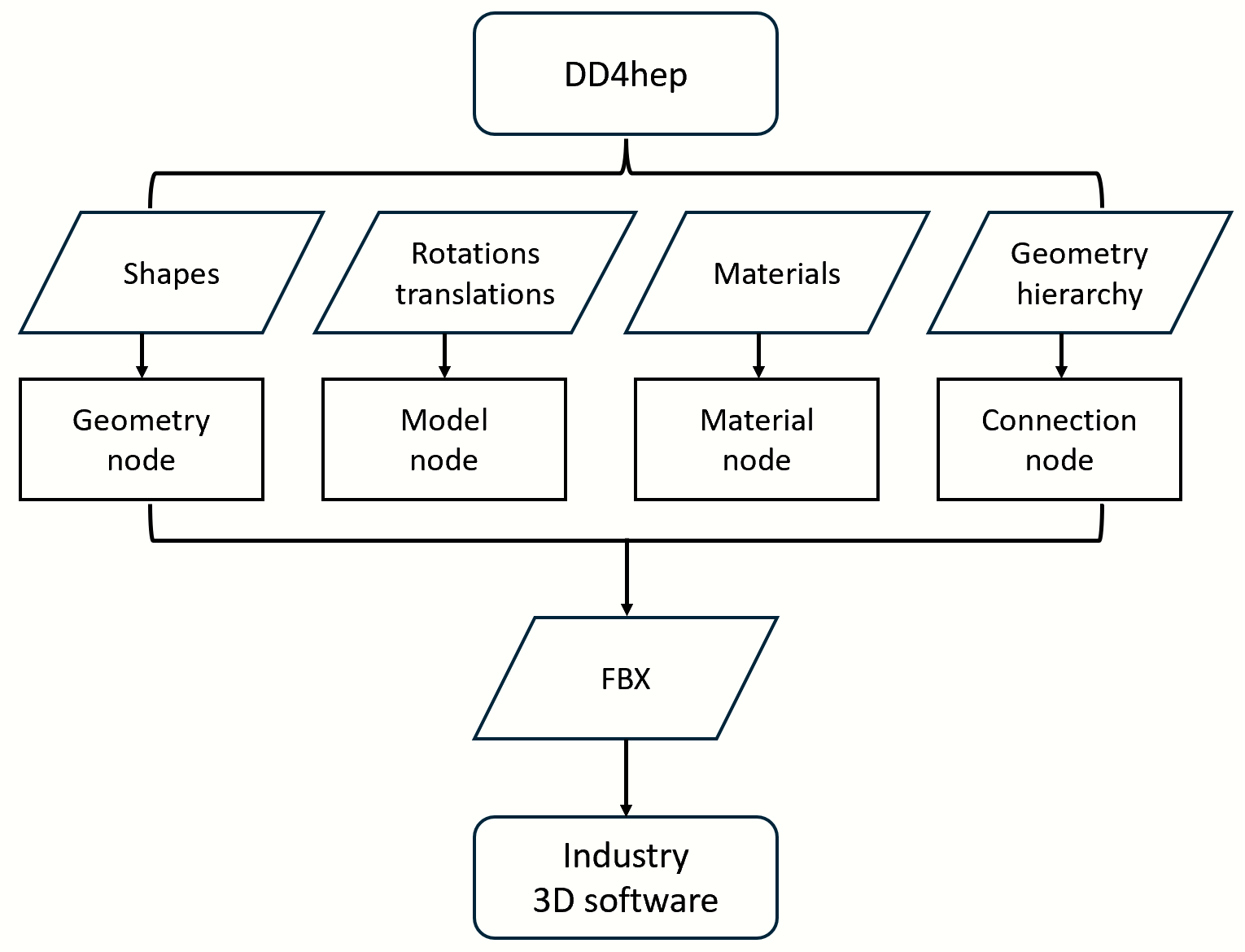}
	\caption{Detector data flow in conversion from DD4hep to FBX.}
	\label{fig:flow}
\end{figure}

To develop a method for converting DD4hep to FBX, the similarities and differences between their detector construction systems should be considered.
As shown in Fig.~\ref{fig:flow}, DD4hep and FBX have four similar components:

\begin{itemize}
\item \emph{Shapes}. 
Dimensions and shapes of the basic solid units that form the detector.
\item \emph{Rotations/translations}. 
Relative positions between different solids.
\item \emph{Materials}. 
Matter, including atom composition, density, and other attributes, from which solids are defined.
\item \emph{Geometry hierarchy}. 
Structures of detector unit compositions and their relationships.
\end{itemize}

To complete the detector data conversion, all geometric information must be extracted from DD4hep, mapped to the corresponding parts in FBX, and reorganized in the format of FBX definition. 
Then, it can be imported into a popular industry software, such as Unity or Unreal, to construct a 3D model of the detectors.

\begin{figure}
    \centering
    \includegraphics[width=0.9\linewidth]{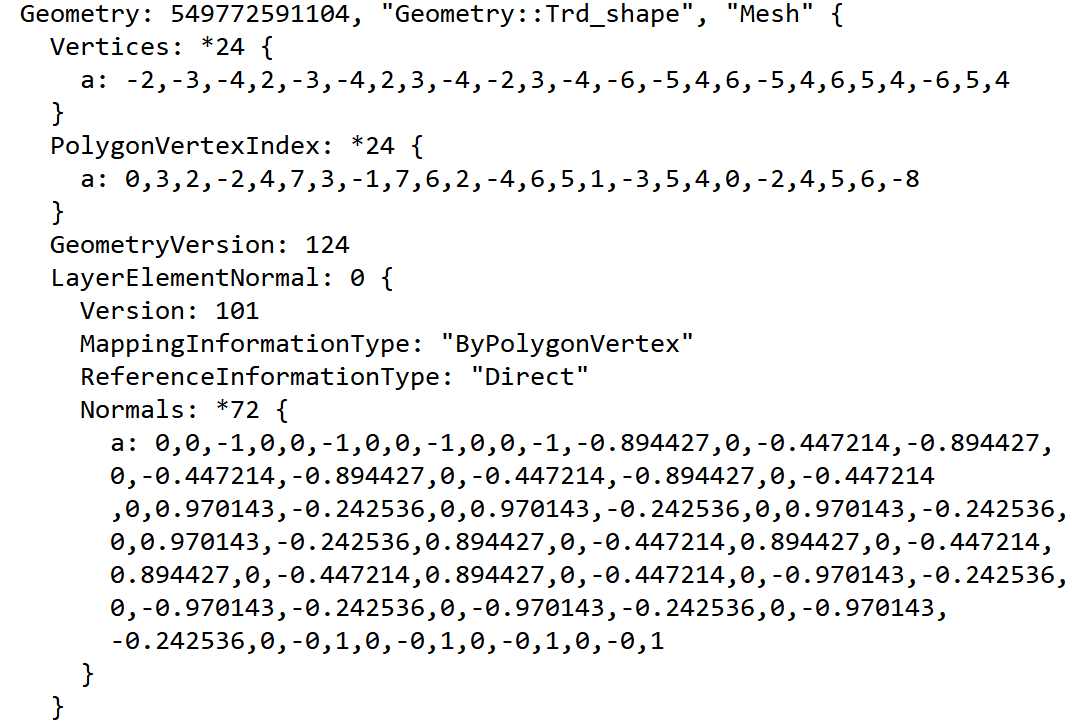}
    \caption{Example FBX file for the Trapezoid shape based on the polygonal surface description.}
    \label{fig:trd_fbx_content}
\end{figure}

The FBX file format is available in an ASCII file format, facilitating reading and checking. 
It comprises a nested list of nodes arranged in a hierarchy. 
In this study, nodes are divided into ``Objects'' and ``Connections.” 
Information regarding shapes, materials, and models, which describe shading, position, and rotation, is placed in the ``Objects'' node. 
The ``Connections'' node describes connections between shapes and models, models and models, or materials and models.
Fig.~\ref{fig:trd_fbx_content} shows a simplified example of the trapezoid ~(TRD2) shape description in FBX. 

\begin{figure}
    \centering
    \subfigure[]{\includegraphics[width=0.43\linewidth]{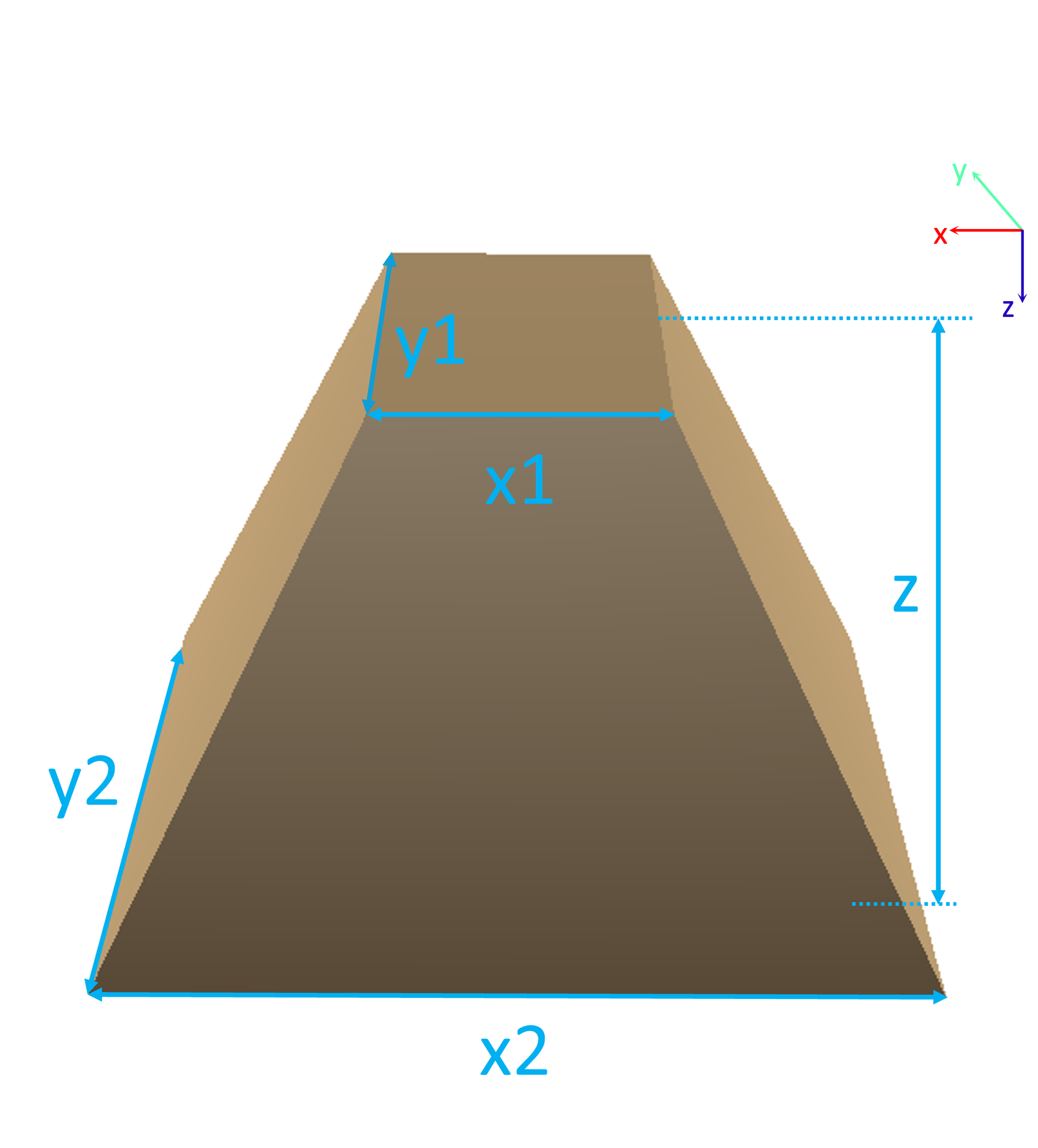}}
    \subfigure[]{\includegraphics[width=0.405\linewidth]{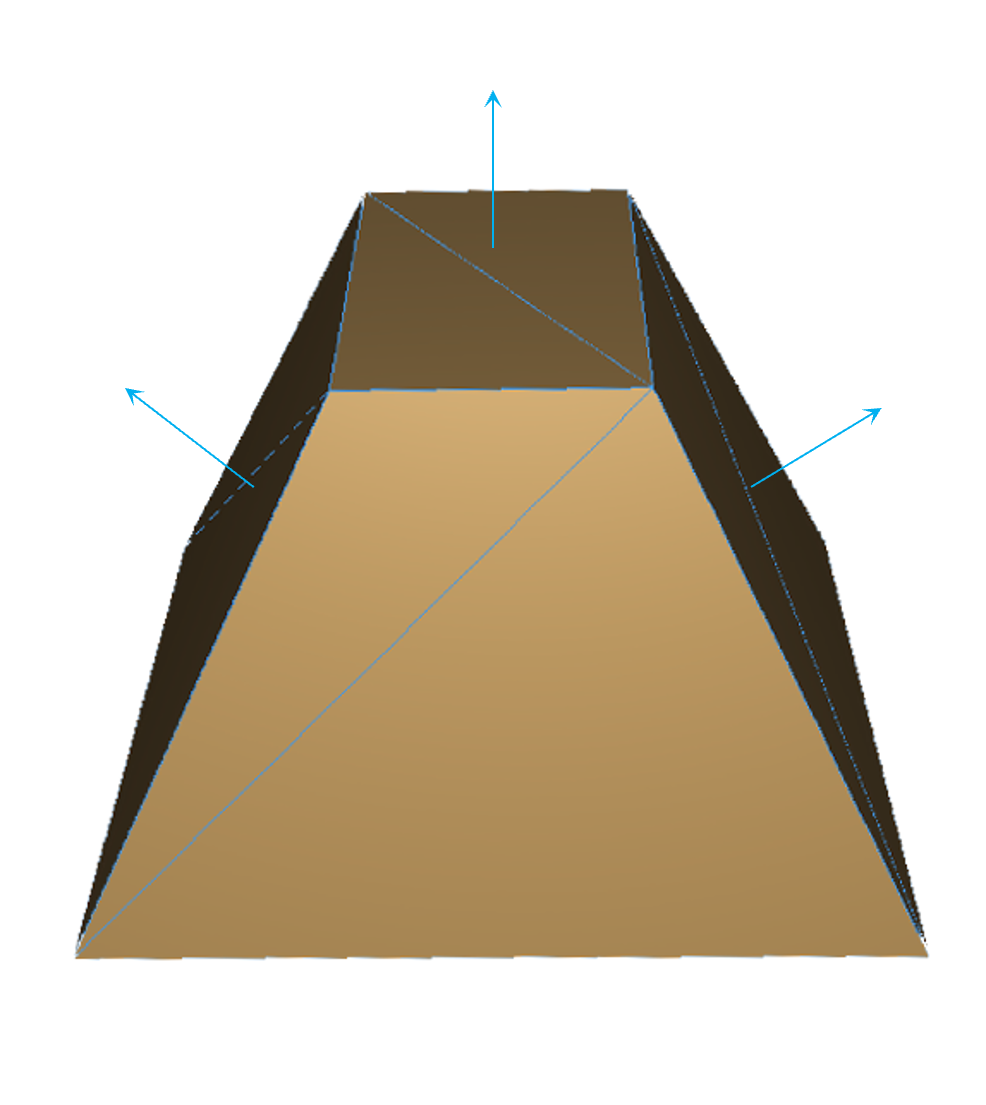}}
    \caption{Visualization of the Trapezoid~(TRD2) shape defined with DD4hep~(left) and FBX~(right).}
    \label{fig:trd_vs}
\end{figure}

The first step in the conversion is to map the shape description into the FBX format for all types of standard shapes in DD4hep.
The current detector description systems in the HEP software, including Geant4, ROOT, GDML, and DD4hep, employ the CSG system, which uses characteristic parameters to define the shape of a solid. 
FBX depends on a surface definition system that uses a set of polygons with vertices and a normal to define the surface of a solid.
For example, as shown in Fig. ~\ref{fig:trd_vs}~(a), to describe a TRD2 shape, which is a trapezoid with both the X and Y dimensions varying along the Z-axis, the CSG system requires only five parameters:
\begin{itemize}
\item \emph{dz}, the half-length along the Z-axis;
\item \emph{dx1}, the half-length along x at the surface at -dz; 
\item \emph{dx2}, the half-length along x at the surface at +dz; 
\item \emph{dy1}, the half-length along y at the surface at -dz; 
\item \emph{dy2}, the half-length along y at the surface at +dz.
\end{itemize}
To describe the same TRD2 shape, FBX needs to determine the coordinates of the eight vertices, the sequence of the vertices to form a polygon, and the normal direction of every polygon to form the surface of the trapezoid, as shown in Fig. ~\ref{fig:trd_vs}~(b).
All this information can be obtained using the five-dimensional parameters of the TRD2 shape definition in DD4hep to generate the corresponding FBX description for TRD2, as shown in Fig. ~\ref{fig:trd_fbx_content}.
Fig. ~\ref{fig:trd_vs} compares the same TRD2 shapes from the DD4hep description visualized in ROOT~(left) and FBX description visualized in Unity~(right).

\begin{figure}[!htb]
	\subfigure[Hyperboloid.]{\includegraphics[width=0.15\textwidth]{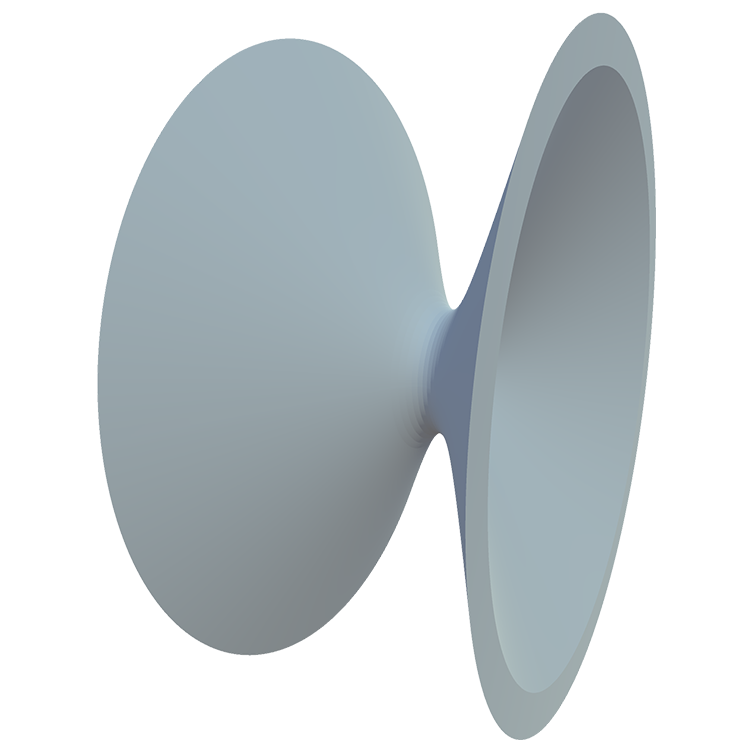}}
	\subfigure[Box.]{\includegraphics[width=0.15\textwidth]{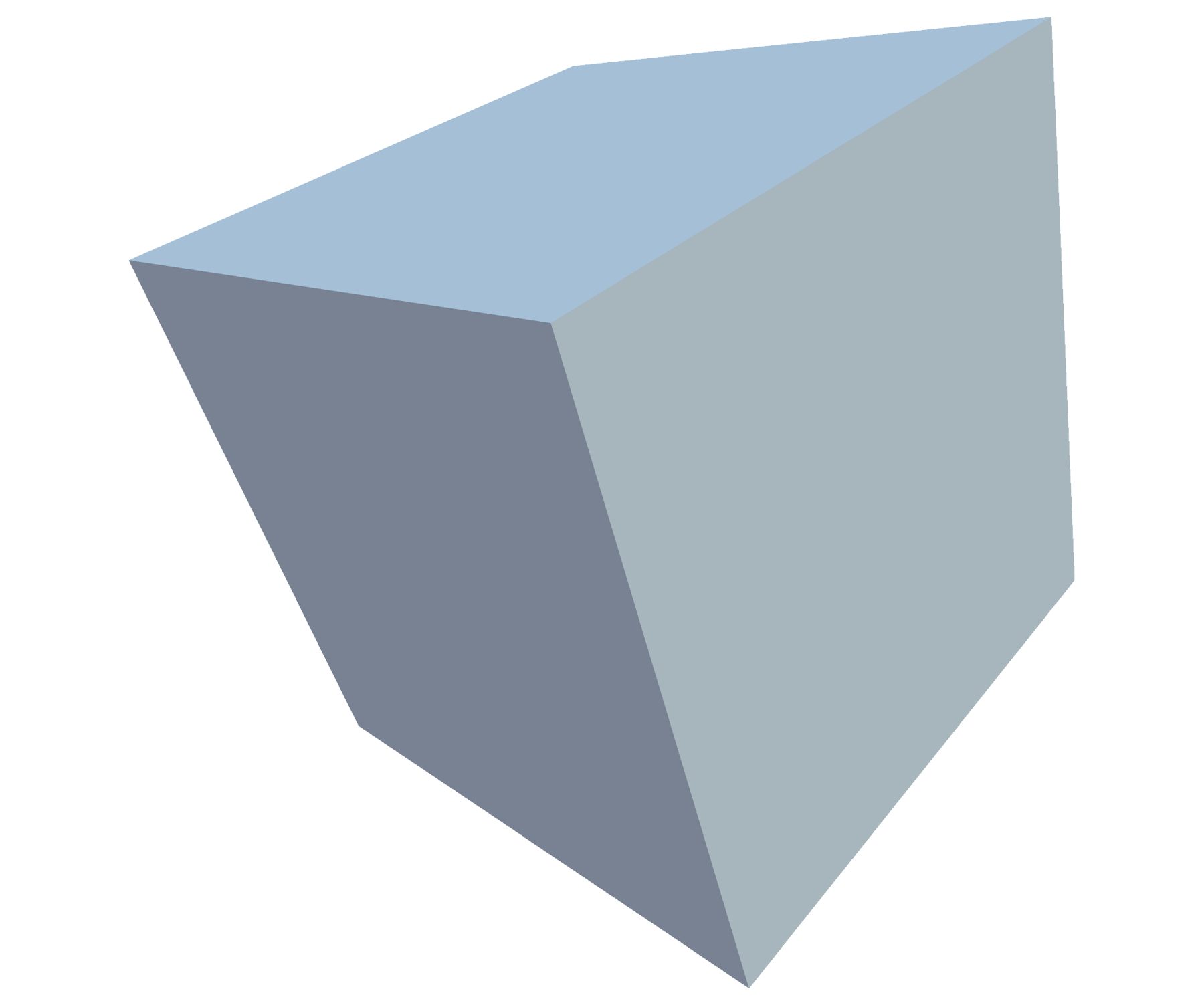}}
	\subfigure[Cone.]{\includegraphics[width=0.15\textwidth]{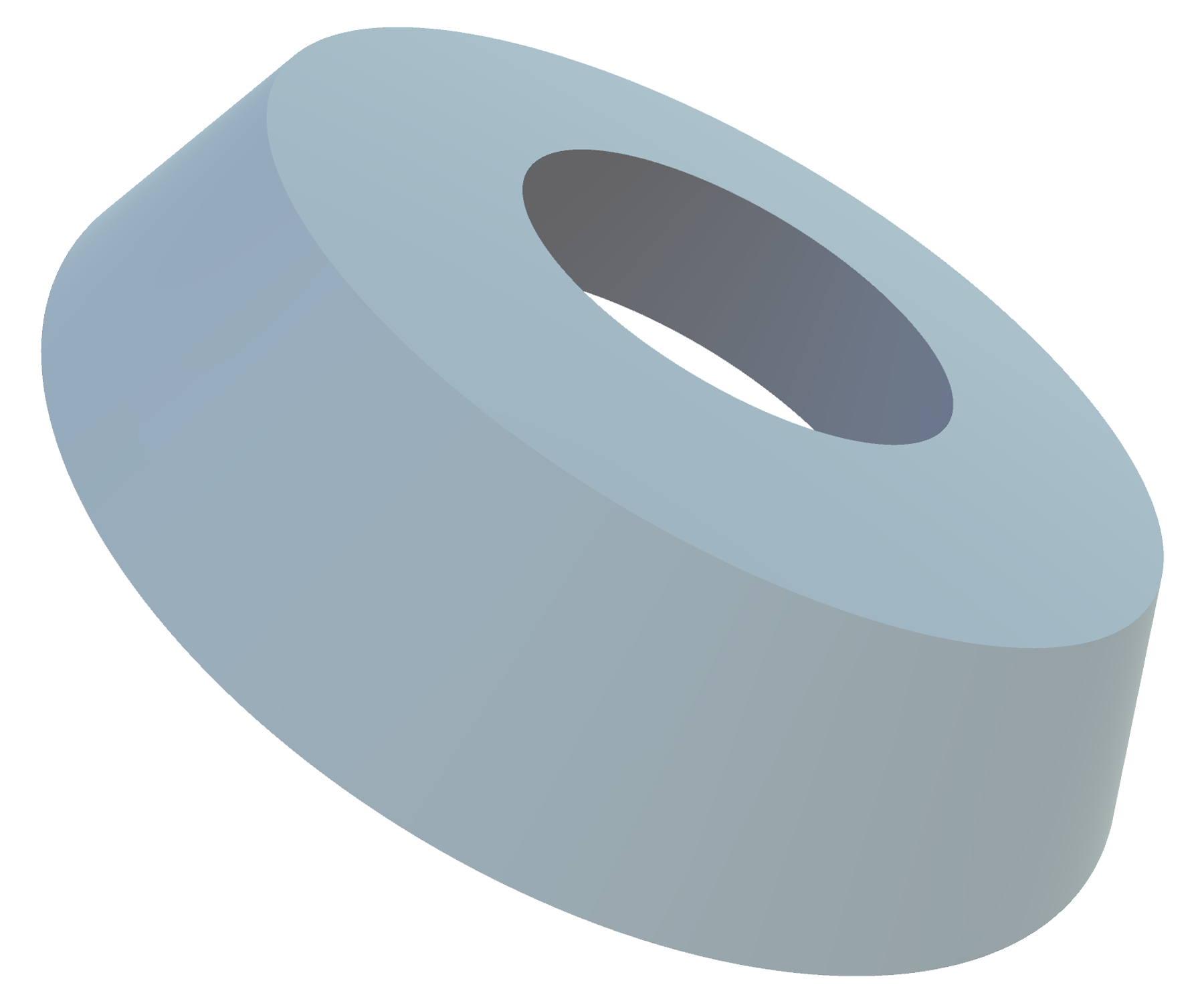}}\\
	\subfigure[Paraboloid.]{\includegraphics[width=0.15\textwidth]{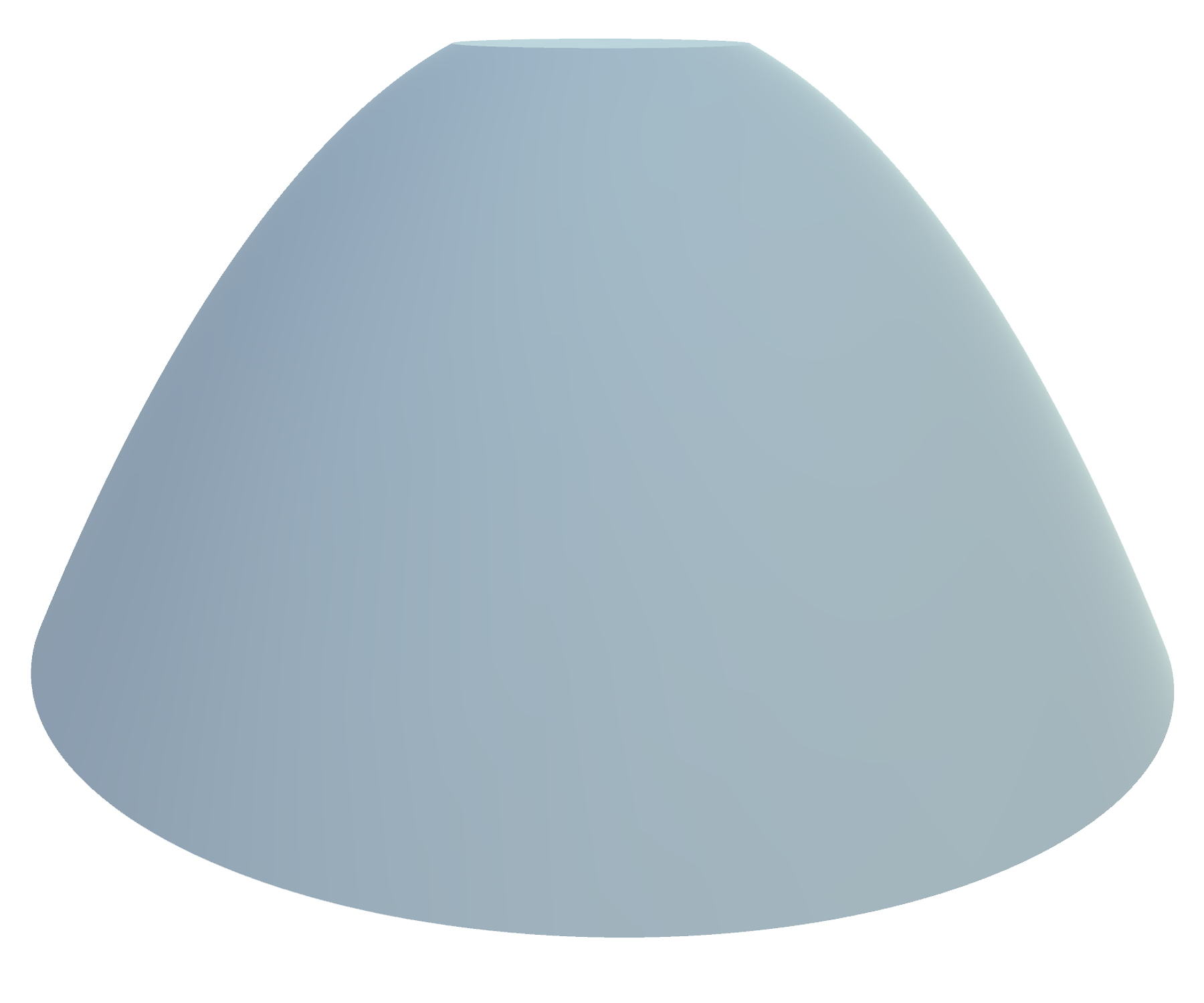}}
	\subfigure[Sphere.]{\includegraphics[width=0.15\textwidth]{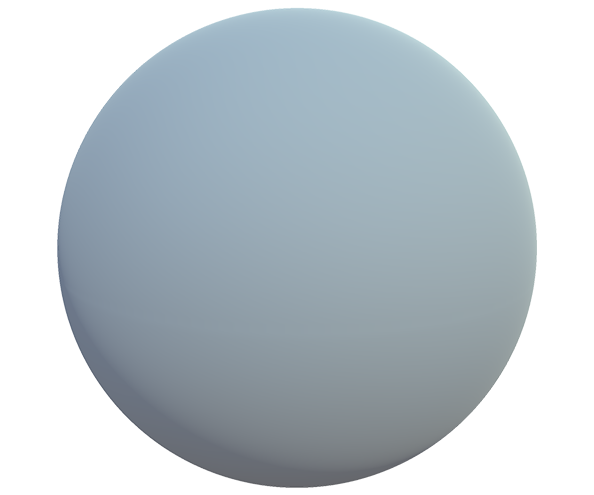}}
	\subfigure[Tube.]{\includegraphics[width=0.15\textwidth]{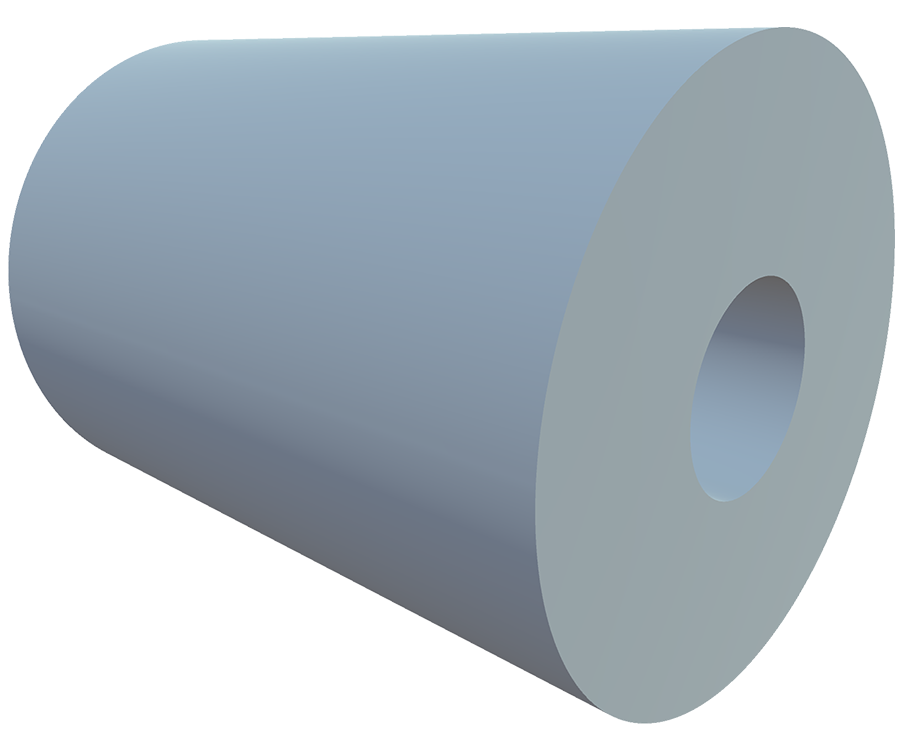}}
	\caption{Some basic shapes converted from DD4hep to FBX and displayed in Unity.}
	\label{fig:baseShape}
\end{figure}

\begin{figure}[!htb]
	\subfigure[Union.]{\includegraphics[width=0.15\textwidth]{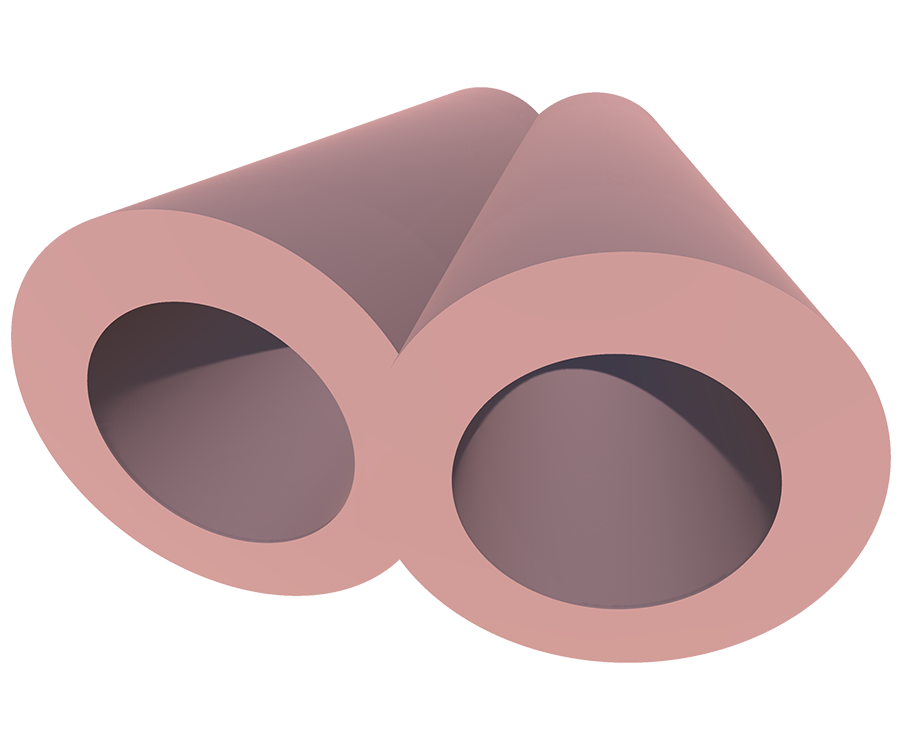}}
	\subfigure[Subtraction.]{\includegraphics[width=0.15\textwidth]{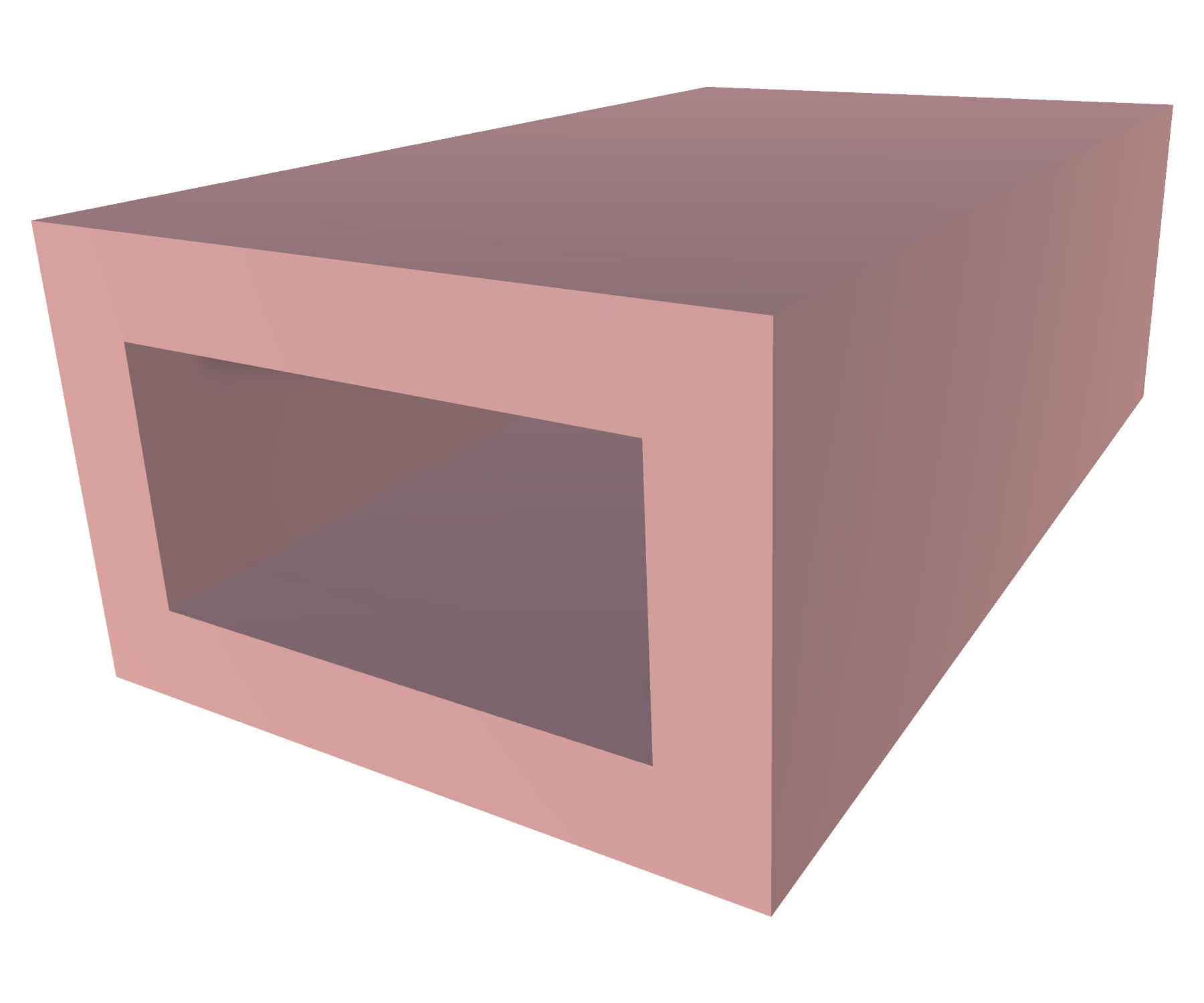}}
	\subfigure[Intersection.]{\includegraphics[width=0.15\textwidth]{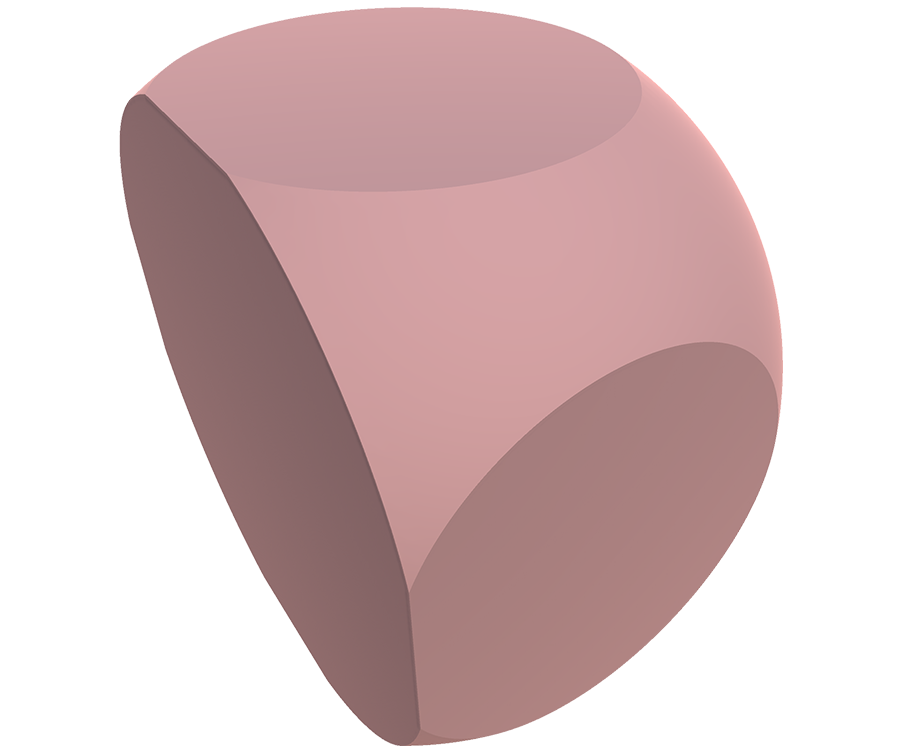}}
	\caption{Boolean shapes converted from DD4hep to FBX and displayed in Unity, including the union of two tubes~(a), the subtraction of two trapezoids~(b), and the intersection of a box and a sphere~(c).}
	\label{fig:compositeShape}
\end{figure}

Similarly, all the other DD4hep shape information can be determined and mapped into the FBX description.
The CSG system is convenient for describing standard shapes, whereas FBX is flexible in describing irregular shapes. 
Fig.~\ref{fig:baseShape} shows standard basic shapes converted from DD4hep to FBX displayed in Unity.
Additional composite shapes are converted in DD4hep using Boolean calculations, including union, subtraction, and intersection, as shown in Fig. ~\ref{fig:compositeShape}.

Second, the transformation information, including translations and rotations of the volumes, in DD4hep is written into the ``Model'' node, which is a sub-node of the ``Geometry'' node in FBX.
The transformation information describes the relative positions for placing a volume within another volume or the ``Mother Volume'' in CSG geometry. 
A volume with such placed information attached is termed as a ``Placed volume'' in DD4hep, ``Physical volume'' in Geant4, or ``Model'' in FBX.
Owing to their similarity, the conversion of the transformation information from DD4hep to FBX is straightforward.

Third, the material information is extracted from DD4hep and written into the ``Material'' node, which is the sub-node of the ``Objects'' node, in FBX. 
In DD4hep, the information of the materials and shapes is attached to the volume, which is also termed as ``Logical volume'' in Geant4. 
Therefore, the corresponding association between materials and volumes must be maintained in FBX. 
In addition to the basic material information from DD4hep, FBX supports attribute definitions, such as texture and reflection, which can be added at a later stage.

Finally, after converting all shape, transformation, and material information, they must be organized in the hierarchy of the entire detector tree with their associations defined in FBX.
A unique index was attached to each geometry, model, and material node. 
Their associations are defined using ``connections,” which are determined by the unique indexes between any two shapes, transformations, materials, and models that are written into the ``connections'' node in FBX.

The original HEP detector description in DD4hep format comprises the following parts: shapes, materials, positions and rotations, detector components (physical node), and detector hierarchy. 
After completing the aforementioned four steps, the DD4hep detector description is transformed into the FBX format, which can be directly read using industrial 3D software, such as Unity and Unreal, while retaining detector information consistent with that in DD4hep.   

Using the aforementioned method, we proposed a method that converted the detector description from DD4hep into the FBX format using automated 3D detector construction. 
The accuracy of the detector geometry conversion in each step was validated by visualization and comparison at both shape and detector levels.   
With the conversion of 3D detector modeling into FBX, further application development using detector geometry in industrial 3D software can be achieved. 

\section{Applications}
\label{sec:application}
 
\subsection{ CLIC detector}

The CLIC is a concept for future linear colliders that can generate $e^+e^-$ collisions at a center-of-mass energy of up to 3~TeV.
Because DD4hep is designed for the next-generation HEP experiments, and the CLIC detector description is already provided in the DD4hep toolkit, the DD4hep--FBX converter should be tested for application in the CLIC detector.

The CLIC experiment requires a detector system with excellent jet energy and track momentum resolution, highly efficient flavor tagging, lepton identification capabilities, full geometrical coverage extending to low polar angles, and timing resolution in the order of nanoseconds to reject beam-induced background. 
The conceptual design of the CLIC detector includes a low-mass all-silicon vertex detector~(VXD) and tracking detector system~(TRD), an electromagnetic calorimeter~(ECAL) with silicon pad sensors and tungsten absorbers, a hadronic calorimeter~(HCAL) with scintillating tiles, a solenoid of 4~T, and a return yoke with an embedded muon detector~(MUD). 

\begin{figure}
    \centering
    \subfigure[]{\includegraphics[width=0.40\linewidth]{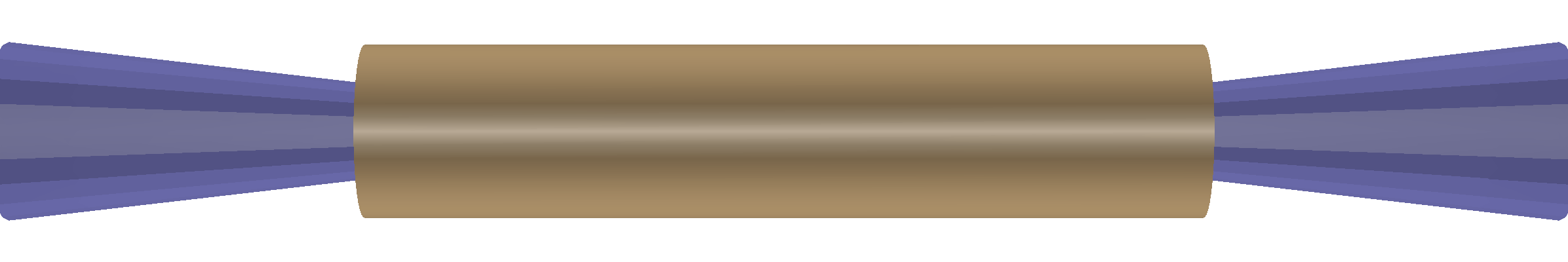}}
    \subfigure[]{\includegraphics[width=0.40\linewidth]{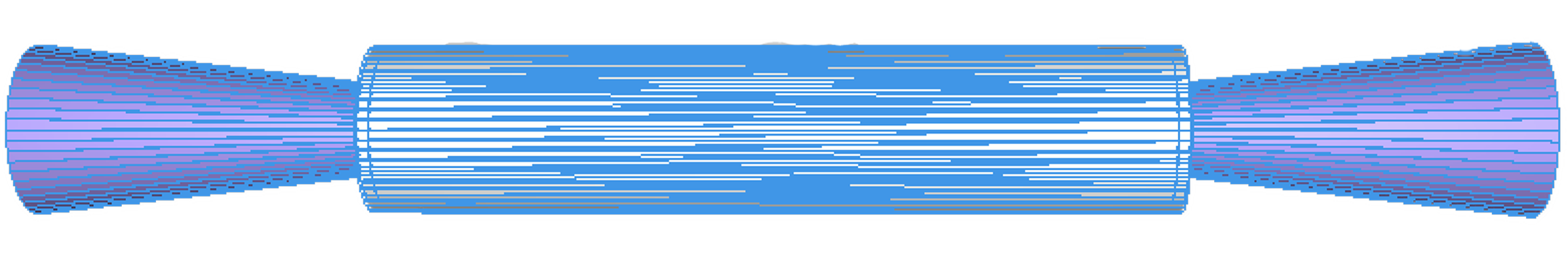}}  \\
    \subfigure[]{\includegraphics[width=0.40\linewidth]{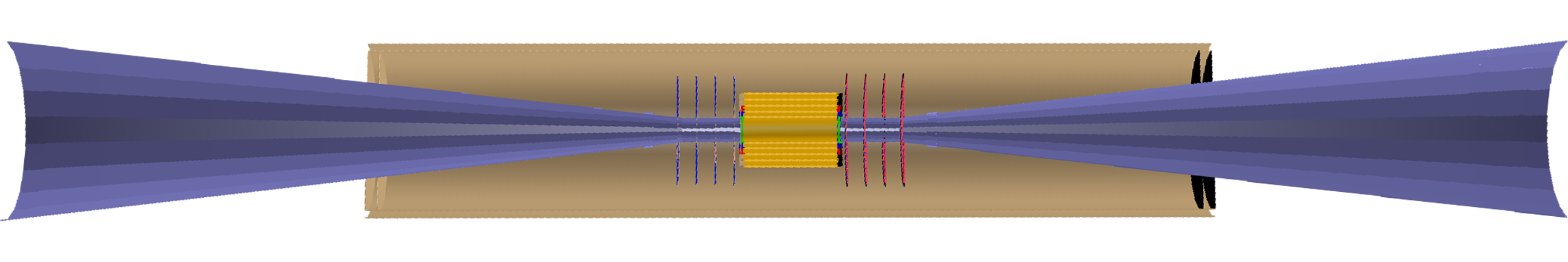}}
    \subfigure[]{\includegraphics[width=0.40\linewidth]{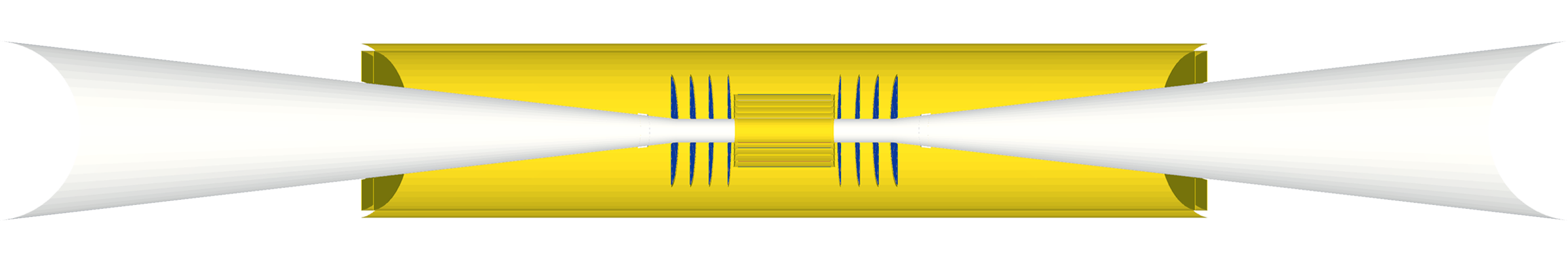}} \\
    \subfigure[]{\includegraphics[width=0.40\linewidth]{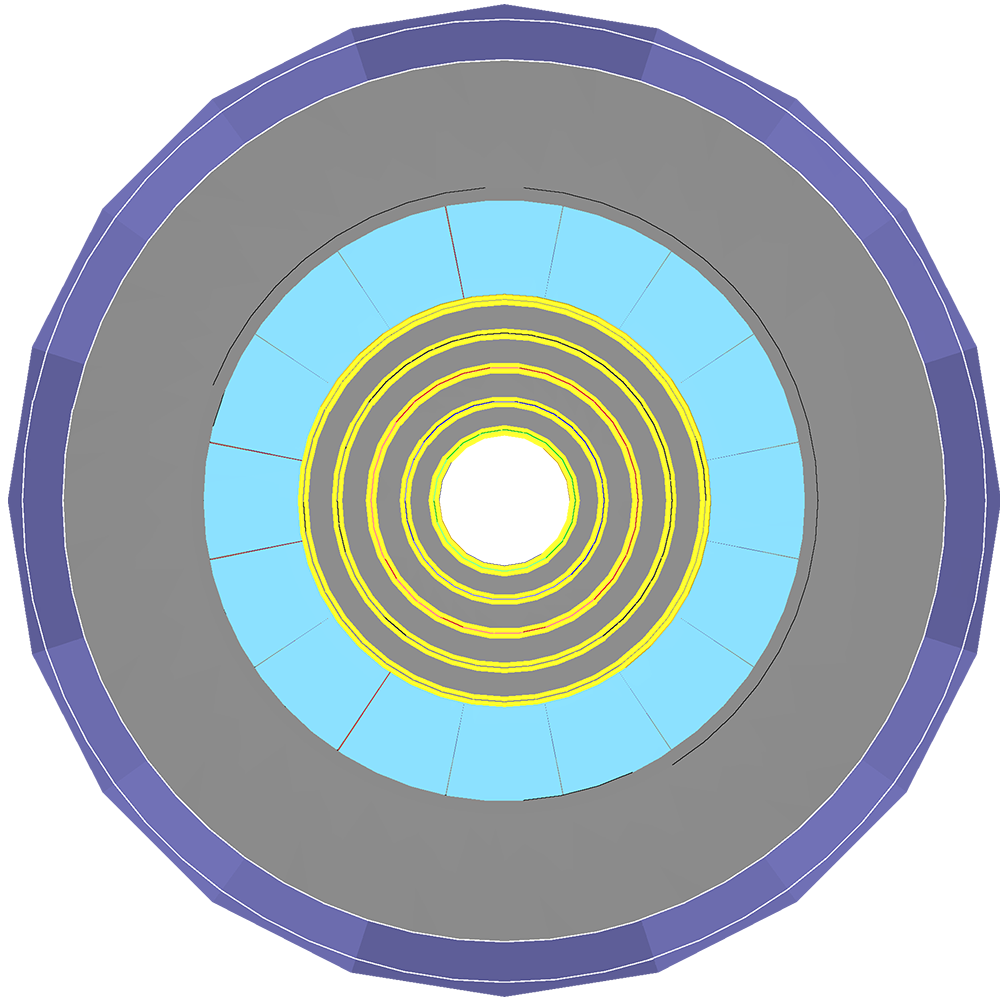}}
    \subfigure[]{\includegraphics[width=0.40\linewidth]{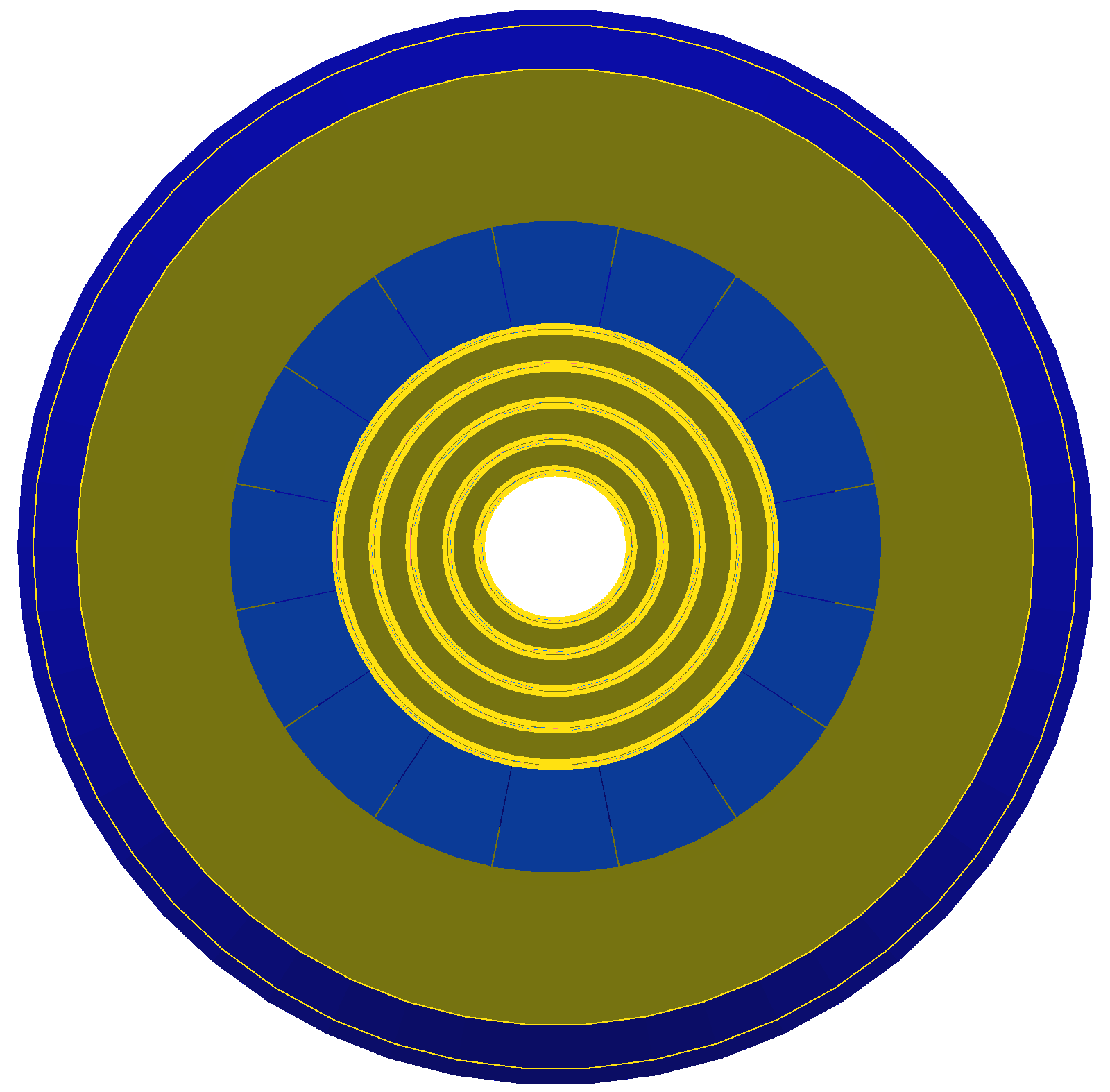}} \\
    \caption{Comparison of the CLIC sub-detector VXD with DD4hep format displayed in ROOT~(left) and with FBX format displayed in Unity~(right). (a)(b) 3D perspective view, (c)(d) cut view of $z-r$ plane, and (e) (f) cut view of $x-y$ plane.}
    \label{fig:clic_vxd_comparision}
\end{figure}

Fig.~\ref{fig:clic_vxd_comparision} shows the CLIC subdetector VXD in the DD4hep format displayed in ROOT~(left) and in FBX format displayed in Unity~(right). 
A 3D perspective view and two cut views of the $z-r$ and $x-y$ planes are compared and validated. 
The silicon VXD consists of five layers in the barrel, eight layers in the endcap, and a carbon fiber support.

\begin{figure}
    \centering
    \subfigure[]{\includegraphics[width=0.37\linewidth]{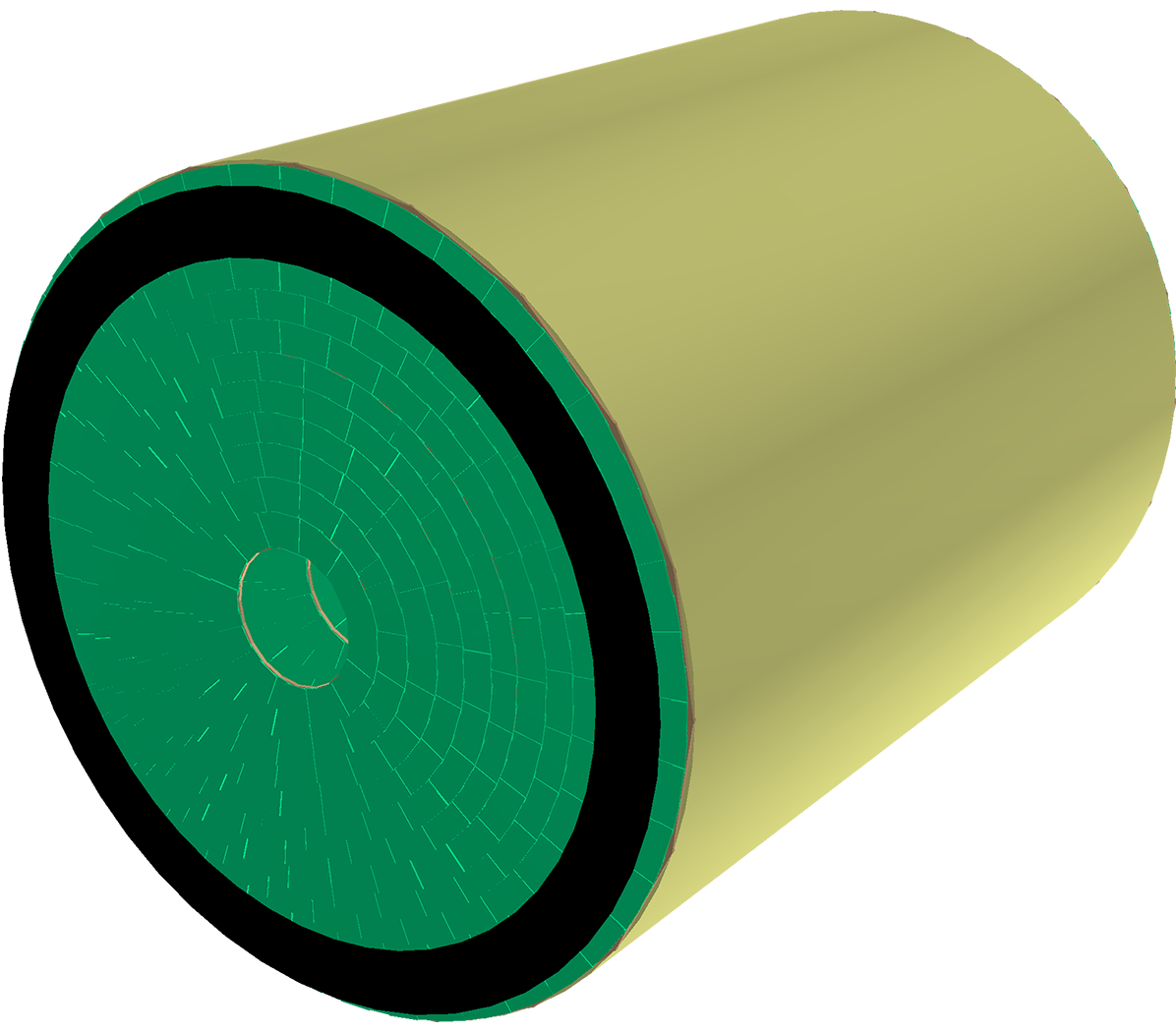} }
    \subfigure[]{\includegraphics[width=0.37\linewidth]{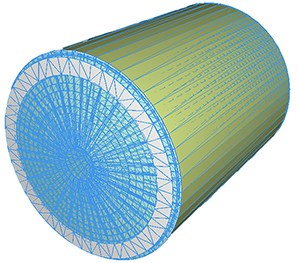}}\\
    \subfigure[]{\includegraphics[width=0.37\linewidth]{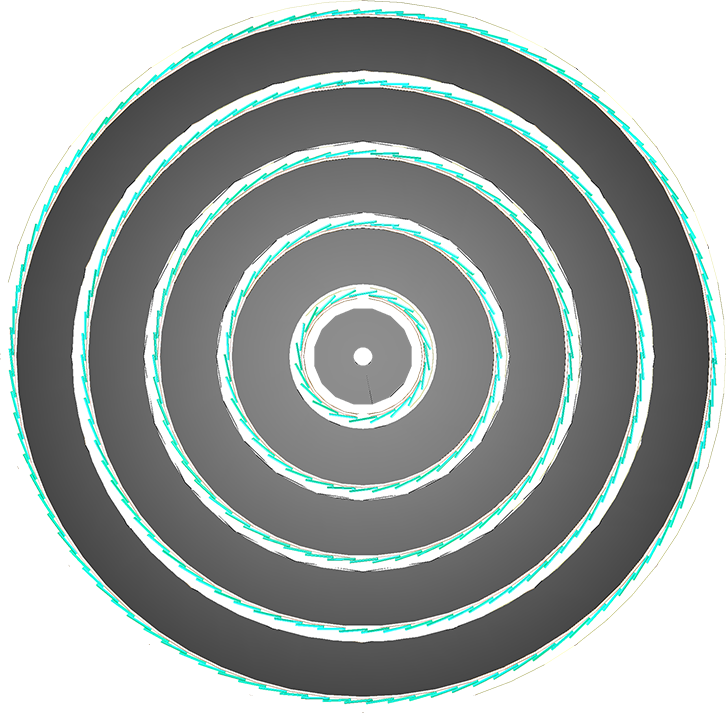}}
    \subfigure[]{\includegraphics[width=0.37\linewidth]{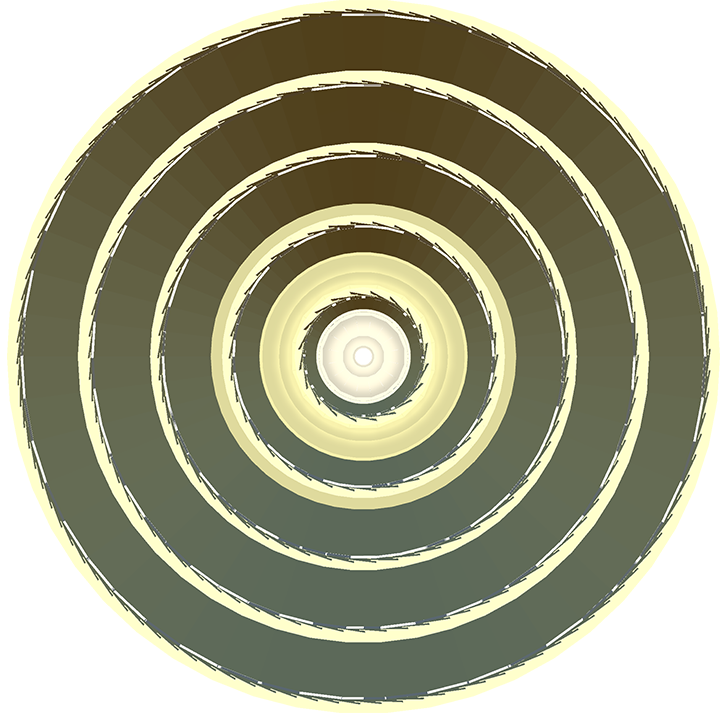} }\\
    \subfigure[]{\includegraphics[width=0.37\linewidth]{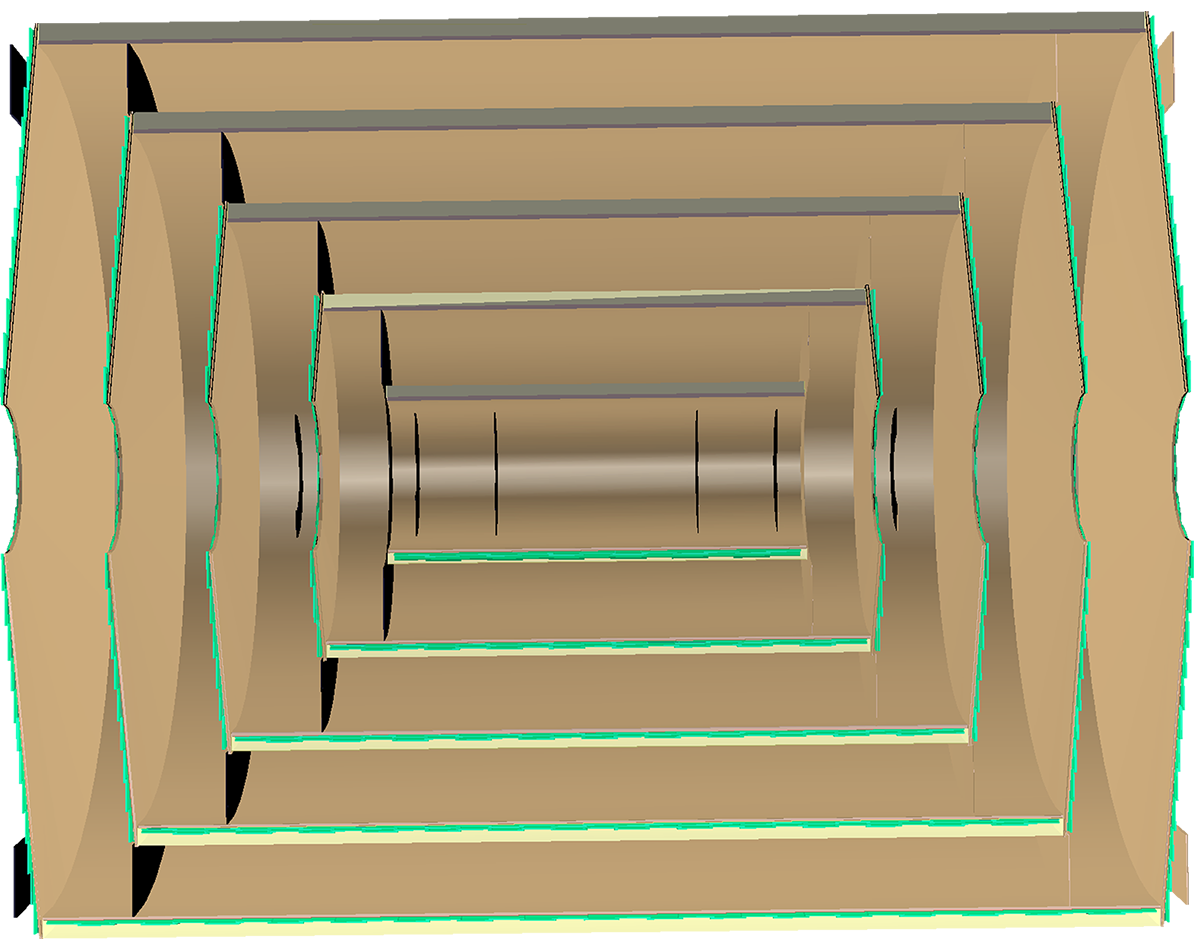}}
    \subfigure[]{\includegraphics[width=0.37\linewidth]{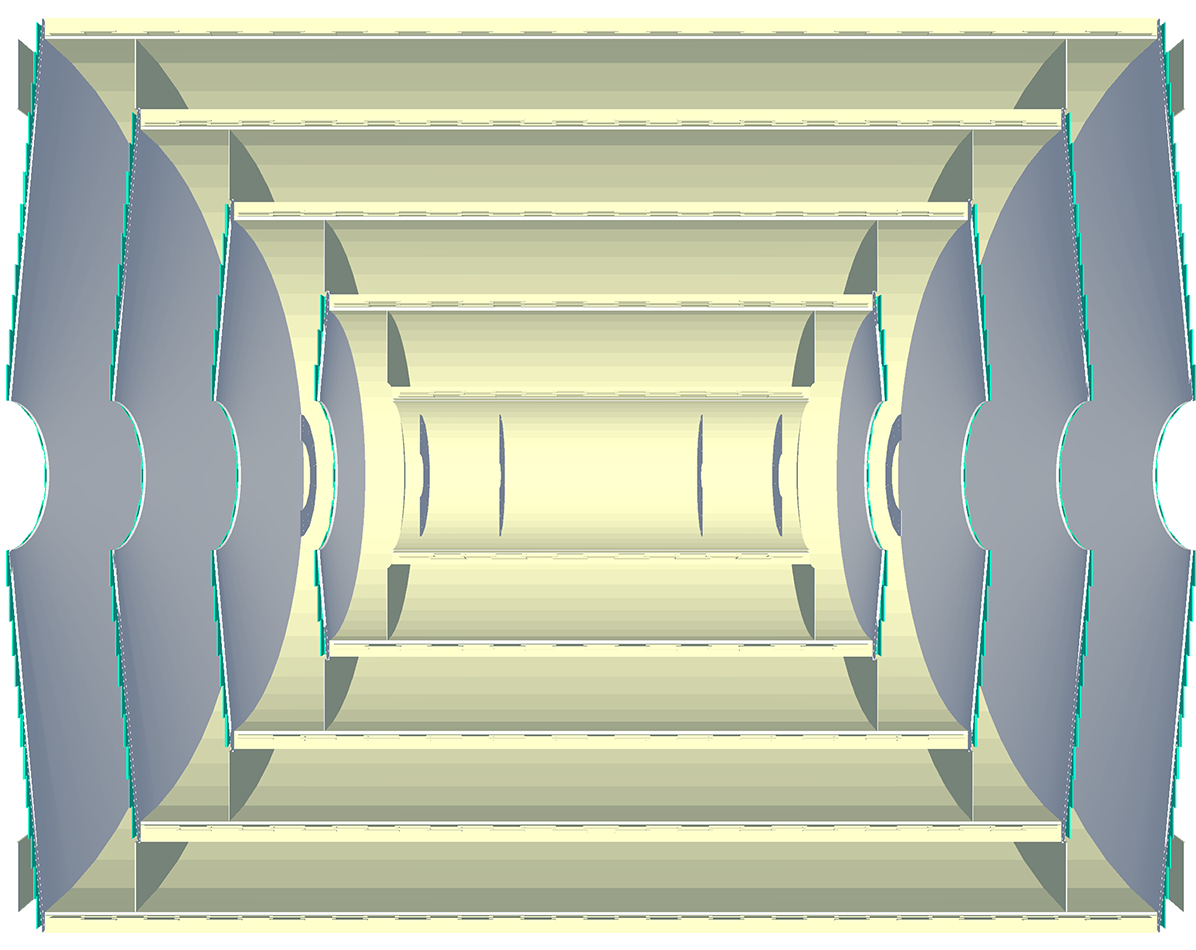}}
    \caption{Comparison of the CLIC sub-detector TRD with DD4hep format displayed in ROOT~(left) and with FBX format display in Unity~(right): (a)(b) 3D perspective view, (c)(d) cut view of $z-r$ plane, and (e) (f) cut view of $x-y$ plane.}
    \label{fig:clic_tracker_comparision}
\end{figure}

Fig.~\ref{fig:clic_tracker_comparision} shows the CLIC sub-detector TRD in both DD4hep and in FBX formats, displayed in ROOT~(left) and Unity~(right)respectively. The TRD consists of five layers in the barrel, eight layers in the endcap, six layers in the forward, and support layers.
Three types of views are also compared.
Because the FBX format uses polygonal surface description, the boundaries of every polygon can be observed in the 3D perspective view of the TRD in this format, as shown in Fig. ~\ref{fig:clic_tracker_comparision}~(b)).

\begin{figure}[htbp]
  \centering
    \subfigure[Solenoid.]{\includegraphics[width=0.21\textwidth]{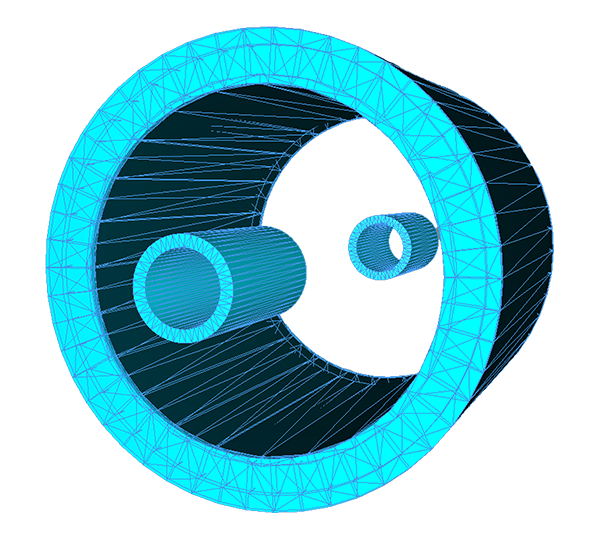}}
    \subfigure[ECAL.]{\includegraphics[width=0.23\textwidth]{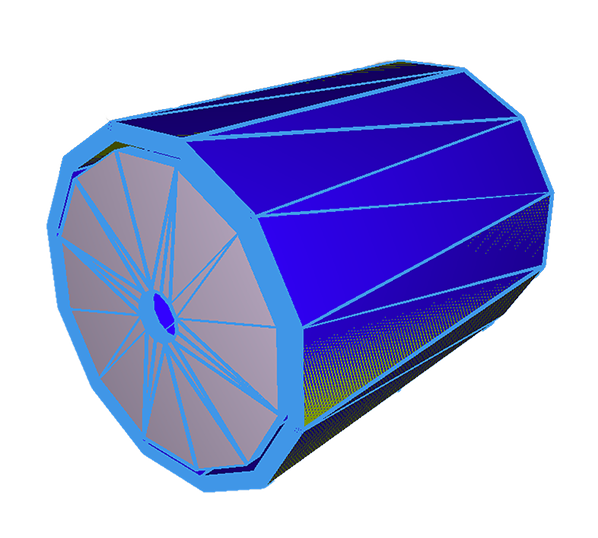}} \\
    \subfigure[HCAL.]{\includegraphics[width=0.23\textwidth]{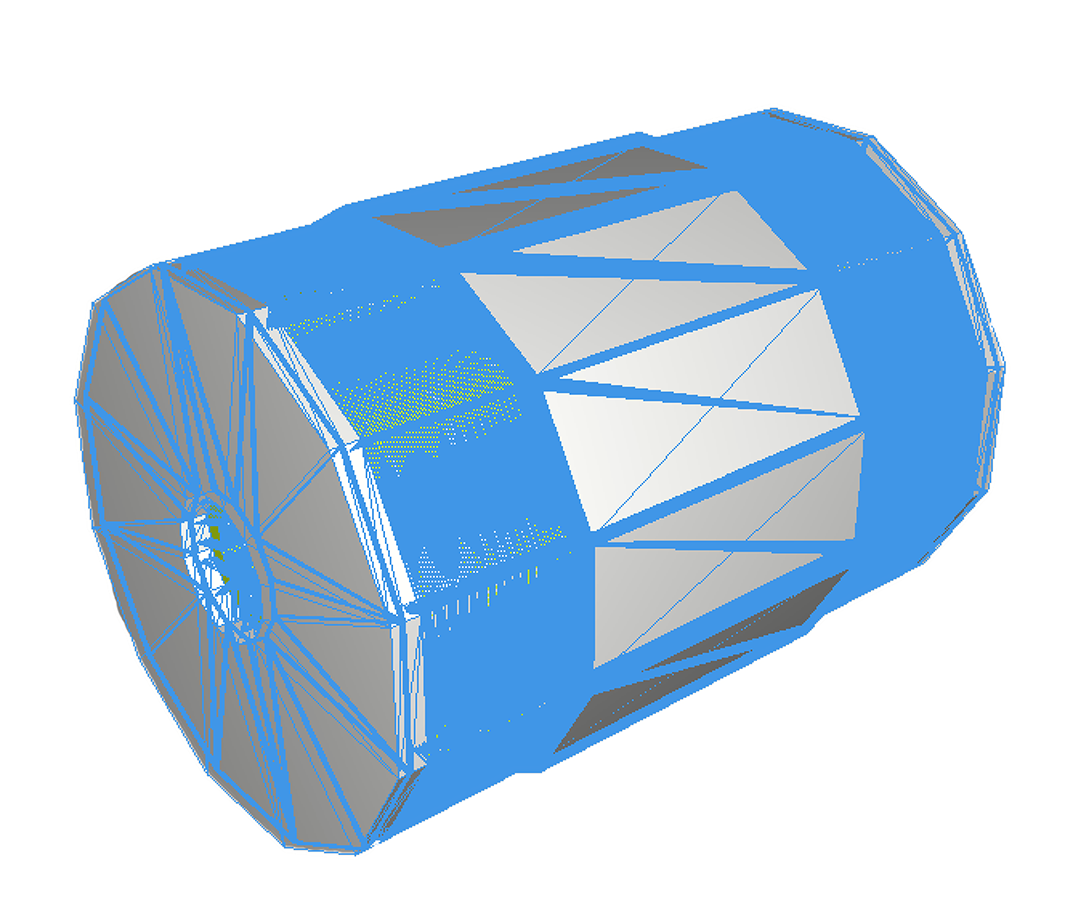}}
    \subfigure[MUD.]{\includegraphics[width=0.20\textwidth]{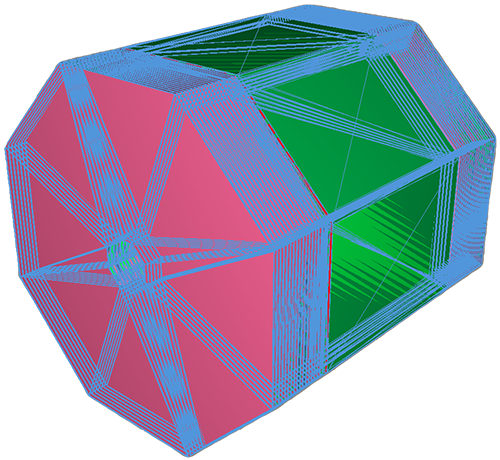}}
    \caption{ CLIC sub-detectors described with FBX and displayed in Autodesk FBX viewer.}
    \label{fig:clic_subDets}
\end{figure}

The other sub-detectors, including the solenoid, ECAL, HCAL, and MUD, of CLIC can be converted from DD4hep to FBX using the same operation.
Their perspective 3D views with FBX descriptions are displayed in Unity, as shown in Fig.~\ref{fig:clic_subDets}.
While displaying the sub-detectors in Unity, the top-world volume and virtual mother volumes should be concealed to enable the inner detector units to be easily identified. 
Moreover, the automatic conversion is advantageous in that the name of each detector unit is maintained throughout the transformation process.
Thus, scripts can set the visualization attributes using the name of each detector unit, rendering it convenient to develop further applications, such as event display. 

\begin{figure}[htbp]
    \centering
    \subfigure[]{\includegraphics[width=0.23\textwidth]{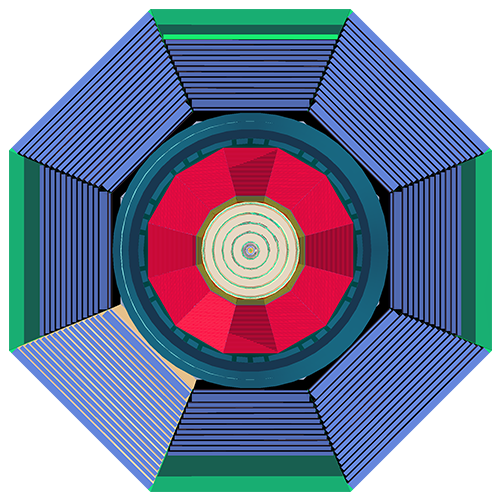}}
    \subfigure[]{\includegraphics[width=0.23\textwidth]{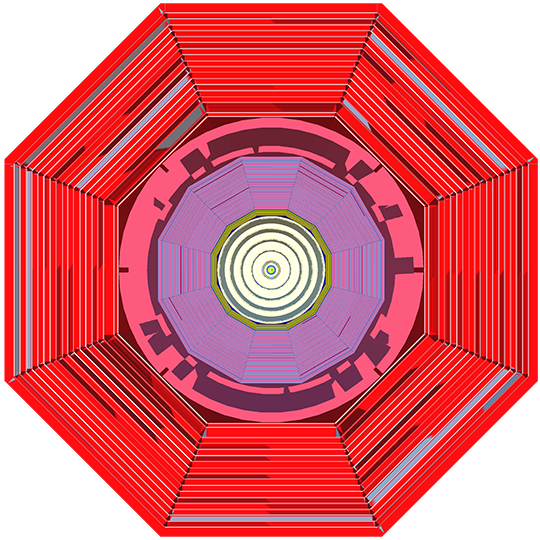}}\\
    \subfigure[]{\includegraphics[width=0.23\textwidth]{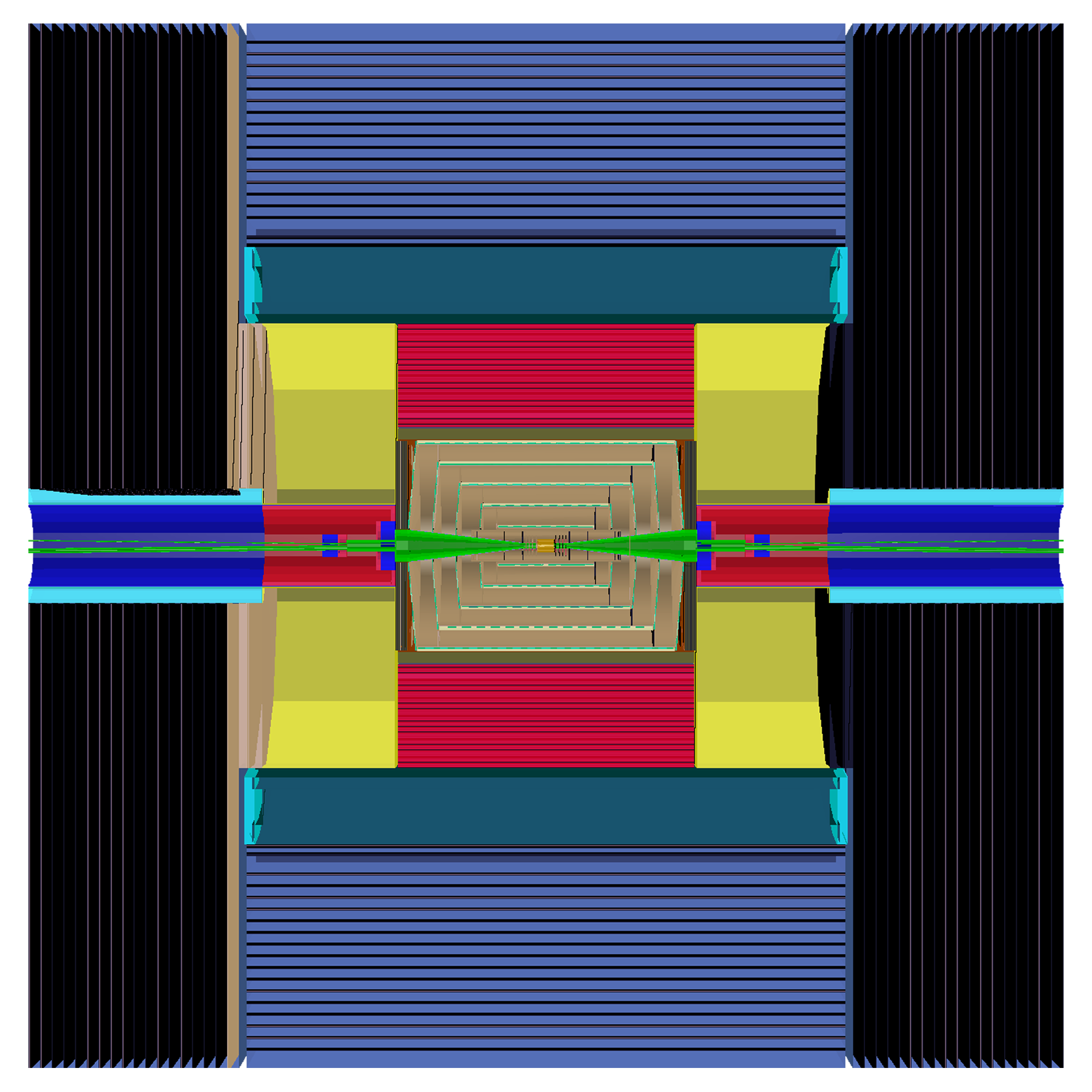}}
    \subfigure[]{\includegraphics[width=0.23\textwidth]{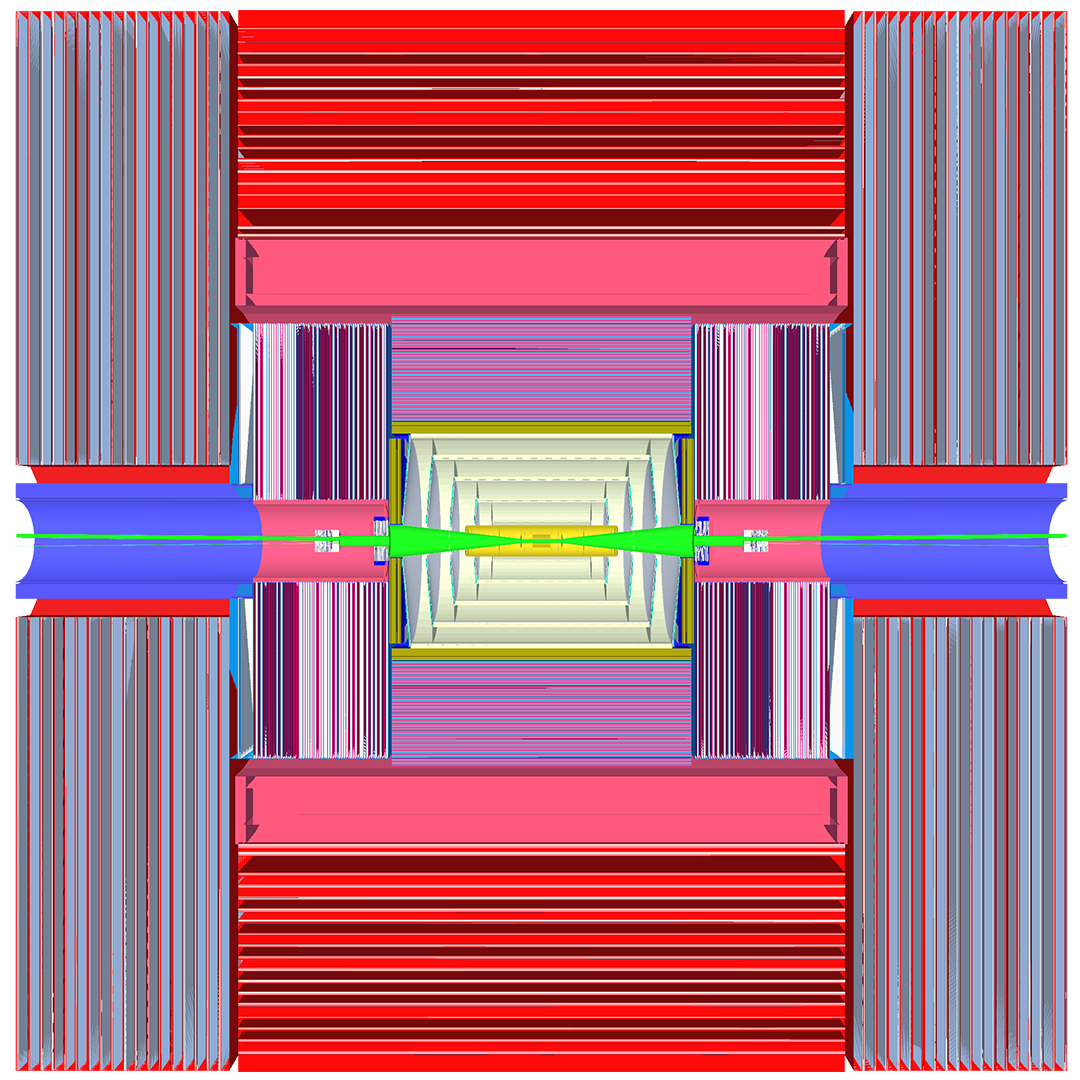}}
    \caption{Comparison of the full CLIC detector with DD4hep format displayed in ROOT (a)(c), and with FBX format displayed in Unity~(b)(d): (a)(b) cut view of the $x-y$ plane and (c) (d) cut view of the $z-r$ plane.}
    \label{fig:clic_fullDet}
\end{figure}

By combining all the sub-detectors, a complete CLIC detector, comprising 
the vertex detector (VTX), TRD, ECAL, HCAL, and MUD sub-detectors from the inside to the outside, are formed.
The display of the full CLIC detector, with the DD4hep format displayed in ROOT and FBX format displayed in Unity, is shown in Fig. ~\ref{fig:clic_fullDet}. 
A converter can also complete the full CLIC detector conversion in a single step and generate a unified FBX file for subsequent development. 
Visual scans of the CLIC detector at both the basic detector unit level and full detector level from various viewpoints reveal no noticeable discrepancies or conflicts between the two formats of the DD4hep and FBX detector descriptions.
To validate the correctness and normal operation of the DD4hep--FBX converter, further applications involving additional detector conversions are necessary to improve its quality. 

\subsection{STCF and CEPC sub-detectors}

The STCF~\cite{STCF_CDR} is an electron-positron collider, which is still in the R\&D stage, with a double ring and symmetric beam energy. It is designed to operate within a center-of-mass energy range of \SI{2}{GeV}$\sim$\SI{7}{GeV} and has a peak luminosity of $0.5\times 10^{35} \rm cm^{-2} s^{-1}$. 

The STCF detector comprises an inner tracker, a main drift chamber, a ring image Cherenkov detector using micro pattern gaseous detector for photon detection, a DIRC-like high-resolution TOF detector~(DTOF), and a crystal ECAL, all of which are enclosed in a superconducting solenoidal magnet providing a 1.0-T magnetic field. 
The solenoid was supported by an octagonal flux-return yoke with resistive plate counter MUDs interleaved with steel. 

Similarly, the CEPC~\cite{CEPC} detector is an electron-positron collider in the R\&D stage. The collider with a circumference of 100 km was designed to operate at a center-of-mass of 240~GeV~(Higgs factory), approximately
91.2~GeV~(Z factory or Z pole), and approximately 160~GeV~(WW threshold scan). 
It contains a vertex detector~(VTX), a silicon inner tracker and silicon external tracker, forward tracking detector and endcap tracking detector components, a time projection chamber, an ECAL, a hadronic calorimeter~(HCAL), a solenoid of 3~Tesla, and a return yoke embedded with MUDs. 

The STCF and CEPC offline software are under development, and the DD4hep toolkit has been used in both experiments for detector description.
Because of the complexity of both detectors and their development stage, we chose only one sub-detector in each experiment to demonstrate the feasibility of applying the DD4hep--FBX converter in STCF and CEPC experiments. 

\begin{figure}[!htb]
    \subfigure[]{\includegraphics[width=0.23\textwidth]{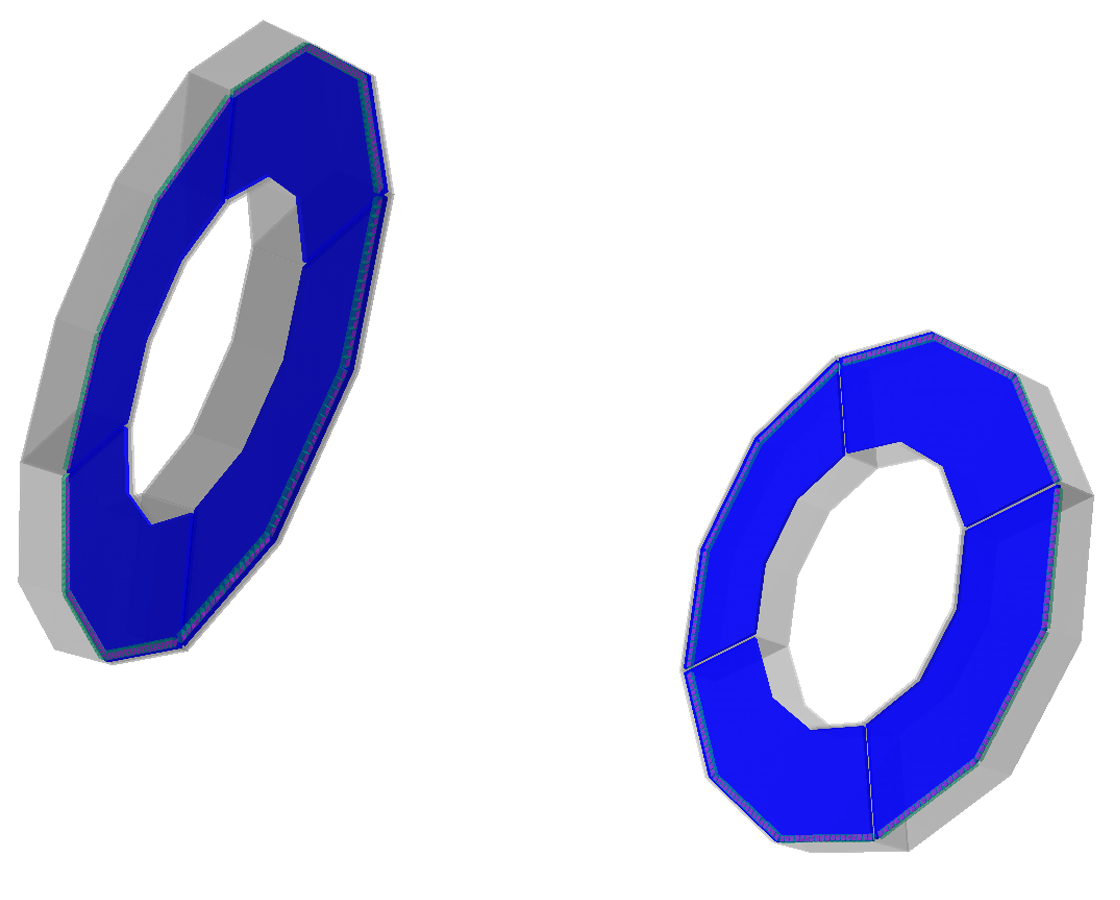}} 
    \subfigure[]{\includegraphics[width=0.23\textwidth]{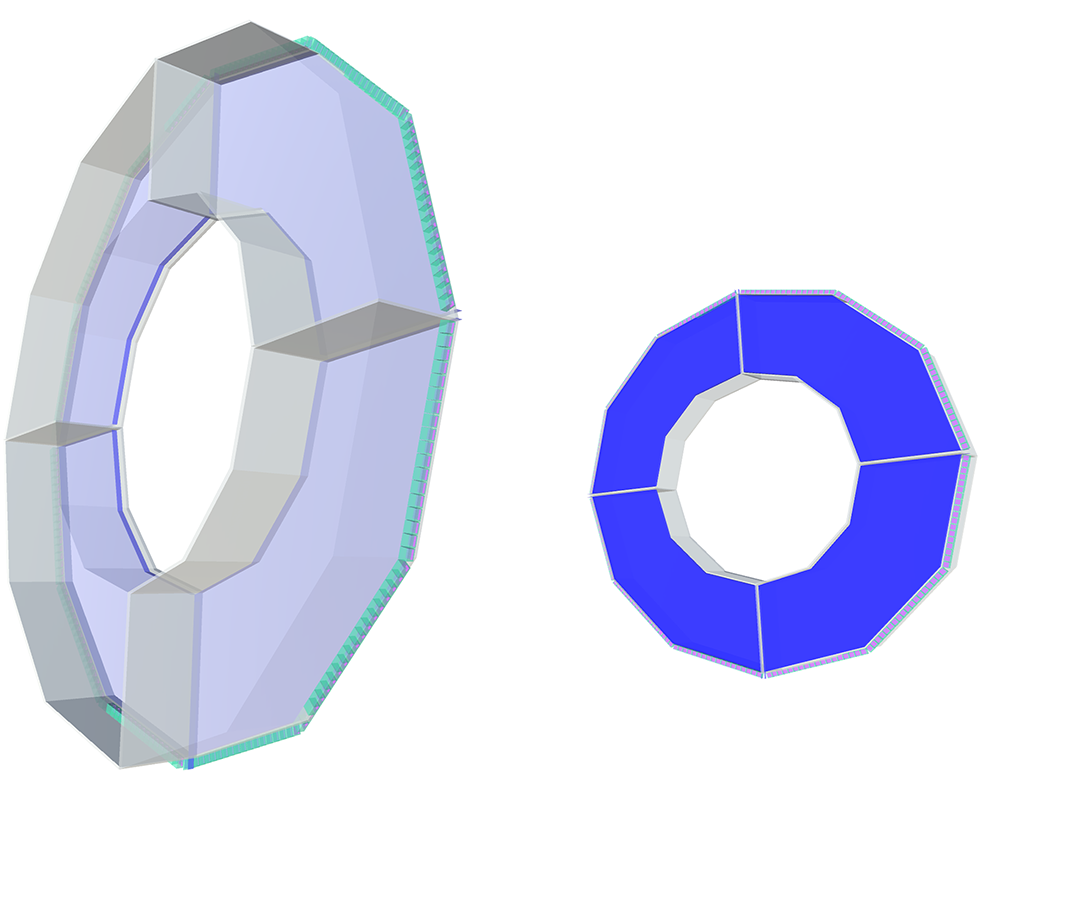}} \\
    \caption{Comparison of the STCF DTOF sub-detector with DD4hep format displayed in ROOT~(a) and with FBX format displayed in Unity~(b).}
    \label{fig:dtofInstcf}
\end{figure}

\begin{figure}[!htb]
    \subfigure[]{\includegraphics[width=0.23\textwidth]{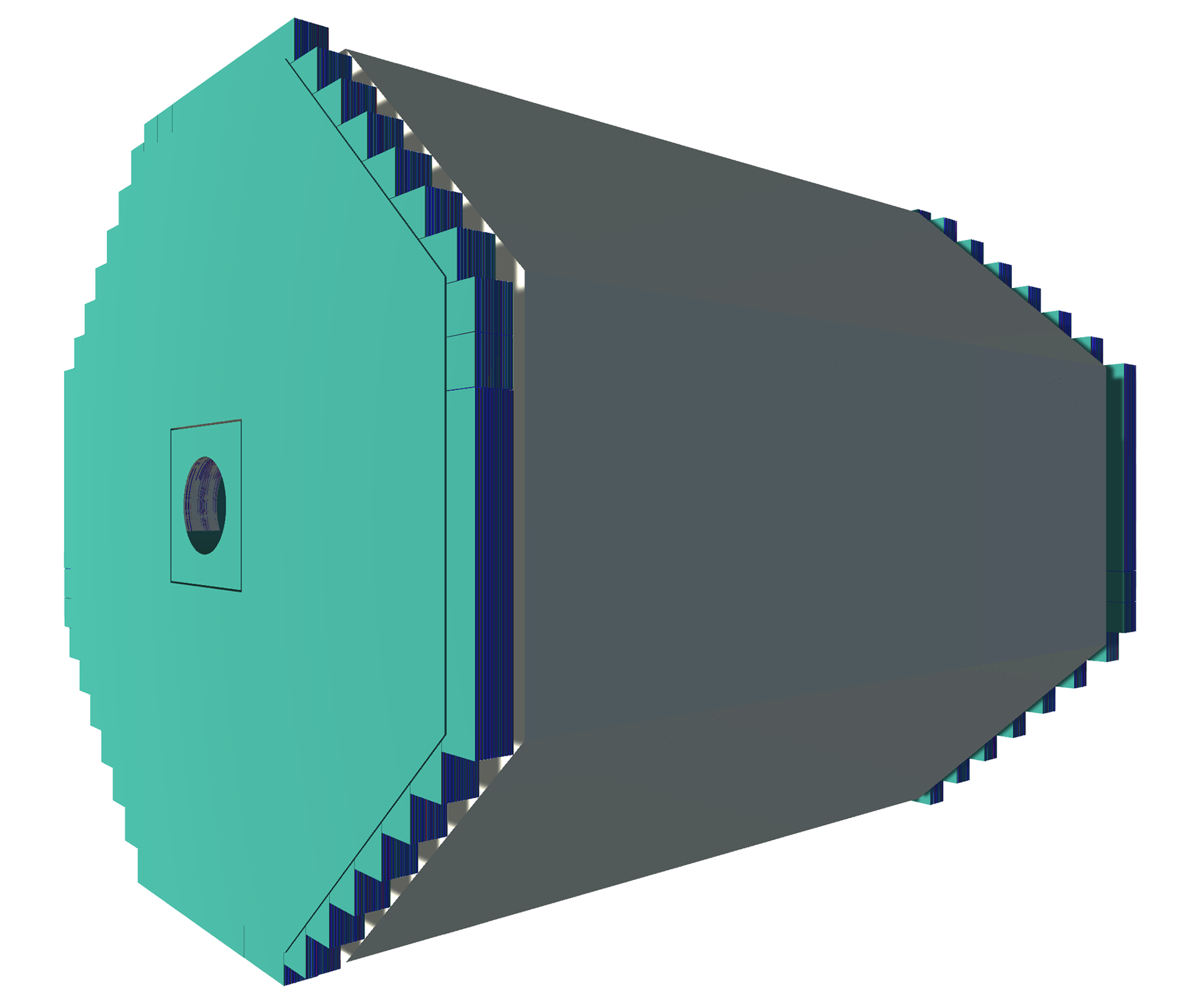}}
    \subfigure[]{\includegraphics[width=0.23\textwidth]{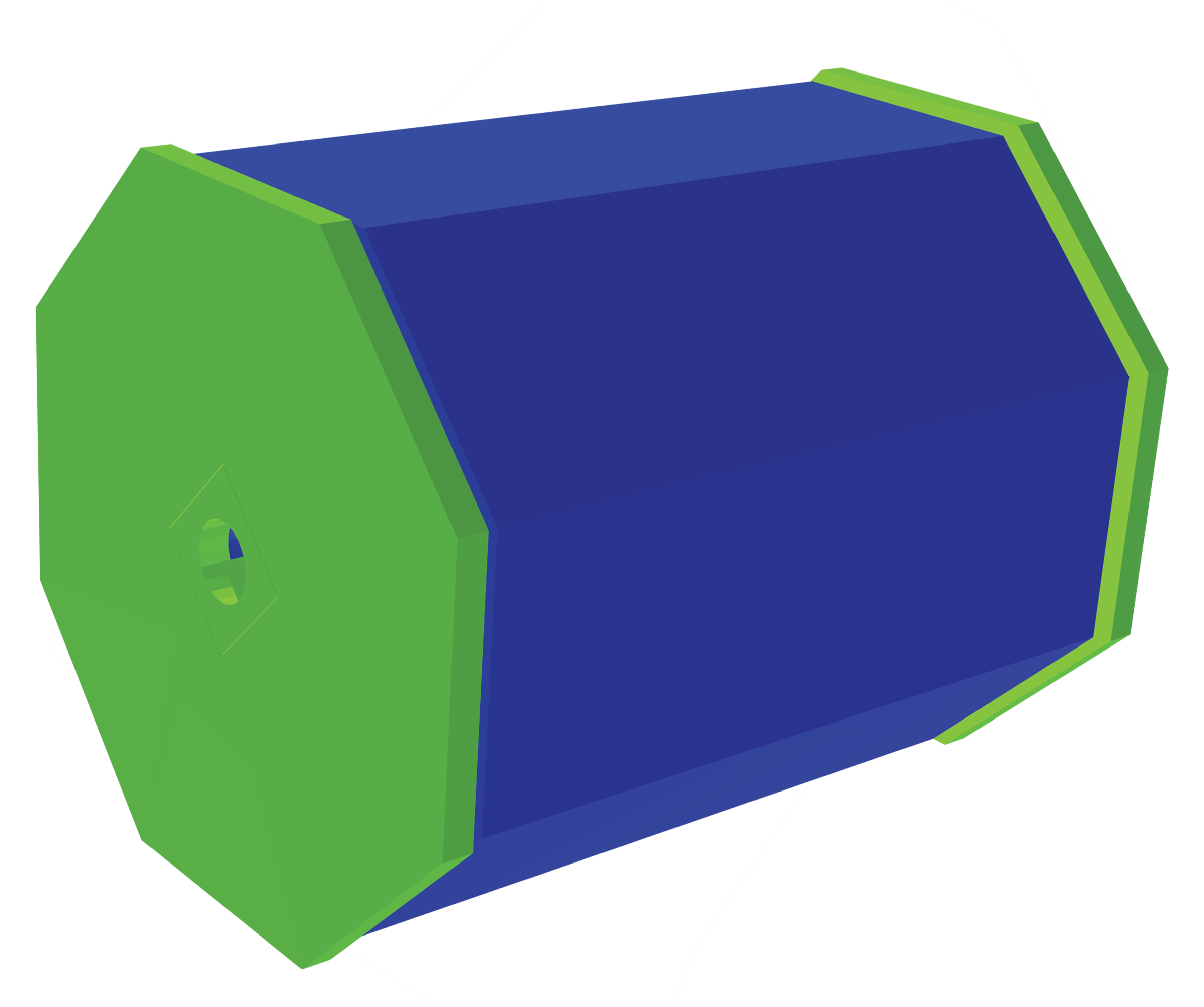}}
    \caption{Comparison of the CEPC ECAL sub-detector with DD4hep format displayed in ROOT~(a) and with FBX format displayed in Unity~(b).}
    \label{fig:ecalIncepc}
\end{figure}

The STCF DTOF and CEPC ECAL sub-detectors were converted from DD4hep to FBX.
The DTOF sub-detector in STCF is a high-resolution detector of internal total-reflected Cherenkov light in the endcap.
A comparison of the DTOF with the DD4hep format displayed in ROOT and FBX format displayed in Unity is shown in Fig. ~\ref{fig:dtofInstcf}.
The ECAL sub-detector in the CEPC is a highly granular silicon--tungsten calorimeter, which can be instrumented to provide precise time measurements with a 50-ps or higher resolution.
A comparison of ECAL in the CEPC with the DD4hep format displayed in ROOT and with FBX format displayed in Unity is shown in Fig. ~\ref{fig:ecalIncepc}.

\subsection{Performance}
\begin{figure}[!htb]
	{\includegraphics[width=0.42\textwidth]{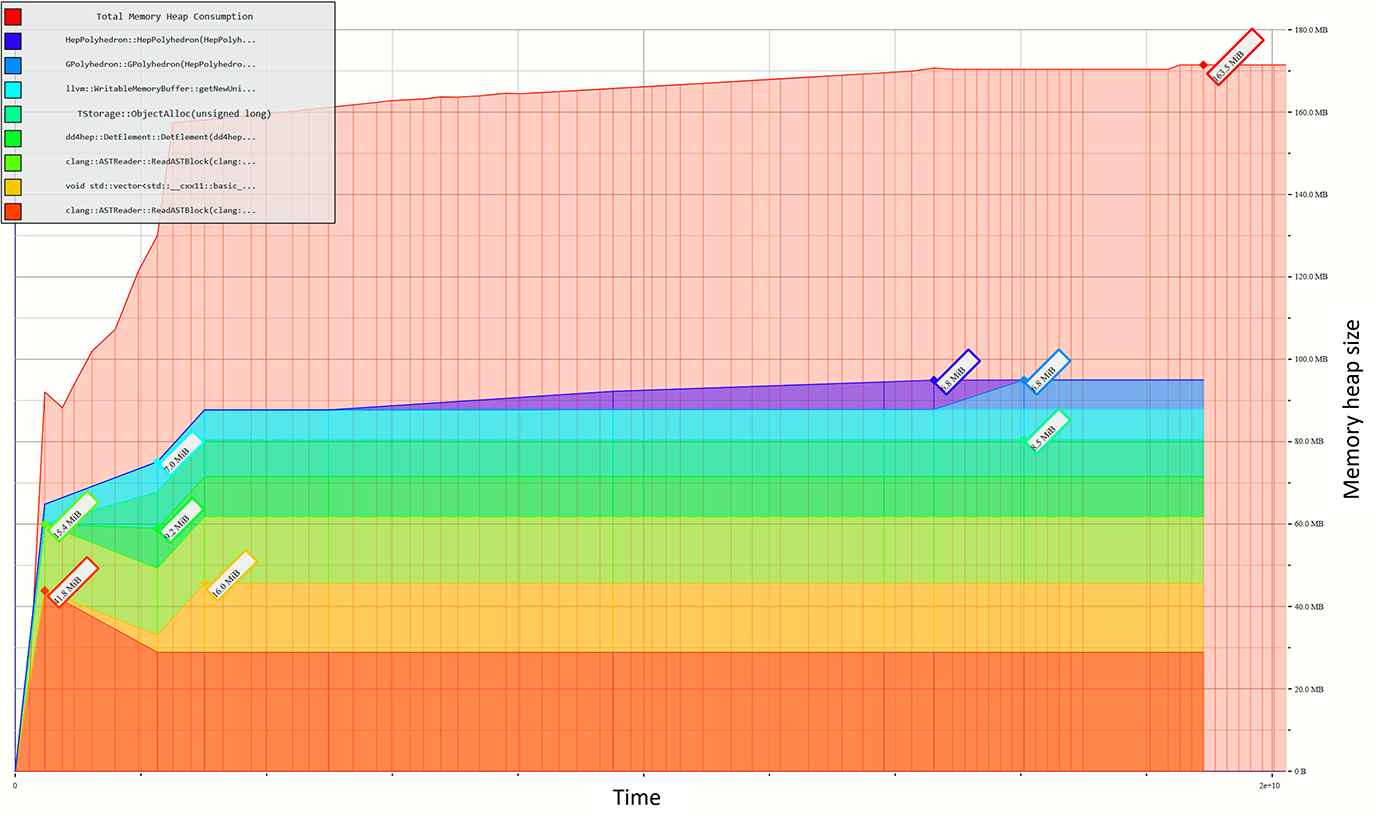}} 
	\caption{Memory usage during conversion of the full CLIC detector from DD4hep to FBX. Purple and blue parts are used to convert shapes to polygons. The cyan part is used to write the FBX nodes to the file on disk. The yellow part is the vector that stores the information of all shapes, volumes, and locations.	}
	\label{fig:memory}
\end{figure}

The performance of the DD4hep--FBX converter was evaluated in the above-mentioned applications.
In the CLIC DD4hep detector, 35,107 detector elements, 11,640 replicated volumes, and 130,882 placed volumes were observed.
The conversion was performed on a computer with an Intel Core i7-13700H CPU~(2.4~GHz) with 16-GB memory.
Approximately 9 s was required to convert the full CLIC detector from DD4hep to FBX, occupying approximately 163.5-MB memory at its peak, using the Valgrind~\cite{nethercote2007valgrind} toolkit for analysis (Fig. ~\ref{fig:memory}).

The total size of the CLIC detector FBX file after conversion from DD4hep was 148.8~MB with 404,306 polygons, which corresponded to the precision of the polygon approximation of the curved surface.
The time and memory consumption for each CLIC sub-detector, as well as the STCF DTOF and CEPC ECAL subdetectors, are summarized and compared in Table~\ref{tab:my_label}.

\begin{table}
    \centering
    \caption{Performance and computing resources consumption in detector conversion.}
    \resizebox{1.0\columnwidth}{!}{
    \begin{tabular}{ccccc}
        \hline
Detector &	Conversion time~(s)	& Memory~(MB) &	Numbers of polygons	& FBX size~(MB) \\
 \hline
CLIC VXD & 0.28	& 92.0	& 22,008	& 1.9 \\
CLIC TRD & 6.07	& 143.9 & 	30,384	& 107.9 \\
CLIC ECAL & 	0.33 & 	93.0 & 	3,300 & 	2.9 \\
CLIC HCAL & 	0.39 & 	94.5 & 	5,100 & 	4.0 \\
CLIC MUD &  0.33 & 	93.0 & 	4,912 & 	3.0 \\
Full CLIC	& 7.46 & 	163.5 & 404,306	& 148.8	\\		
STCF DTOF & 0.49 & 	92.7 & 	188	& 6.4 \\
CEPC ECAL &  28.28 & 358.5  &  46,526 & 459.0 \\  
         \hline
    \end{tabular}
    \label{tab:my_label}
}
\end{table}

Because of the complex CLIC full-detector geometry and large polygon numbers in FBX, more than 2 h is required to import the CLIC FBX file into Unity.
Fortunately, the import process from FBX to Unity is a one-time operation. 
After importing the FBX file and saving it as a Unity project, approximately 30 s is required to reopen the project. 
Nevertheless, further improvements are required for developing the DD4hep--FBX or similar converters in the future.

\subsection{Further application development}
\label{sec:further}

With the HEP detector description converted into the FBX format, FBX files can be directly read by the industrial 3D software to further develop applications, such as event display tools, VR or AR programs with excellent visual effects, and cross-platform deployment.

Unity is a well-known video and game production engine.
Recently, it has been used in HEP experiments, such as the Cross-platform Atlas Multimedia Educational Lab for Interactive Analysis (CAMELIA) ~\cite{CAMELIA,CAMELIA_web} in the ATLAS experiment and Event Live Animation with Unity for Neutrino Analysis (ELAINA)~\cite{ELAINA} in the JUNO experiment, for event displays.
In both programs, the detectors were manually constructed in Unity, rendering them one of the most time-consuming parts of the program development.
With the DD4hep--FBX converter, detector construction in Unity will be convenient for future HEP experiments.

Unreal Engine is another potential and versatile industrial software that can create 3D content for applications such as games, films, TV, architecture, and automobiles.
Both Unity and Unreal are potential tools for developing VR and AR applications with strong hardware support.
Some VR applications have been realized in currently running HEP experiments, such as ATLASrift~\cite{ATLASrift}, BelleII VR~\cite{BelleIIVR}, and VR tours for experiments at CERN~\cite{CERN_VR}.
VR and AR programs provide sufficient experience for understanding HEP detectors and experiments, significantly contributing to detector design, data analysis, and outreach.
The DD4hep--FBX converter can assist and accelerate the development of these programs owing to its automatic and fast detector conversion function.

\section{summary}
\label{sec:summary}

In the offline software development of the next-generation HEP experiments, the HEP software community has adopted DD4hep as a common technique for detector description. However, in contrast to commonly used 3D modeling formats such as FBX, the DD4hep format cannot be directly imported by most well-known 3D modeling and visualization software from the industry. Consequently, this potential industrial software cannot easily be used for application developments in HEP experiments. 

To address this limitation, we introduced a novel method for converting detector description from DD4hep to the FBX format. A full CLIC detector and several sub-detectors of the STCF and CEPC were tested to demonstrate the feasibility of this method. Using the automatic conversion interface, we provided convenience and potential for leveraging industrial 3D software in developing detector geometry and visualization-related applications, such as detector design, simulation, event display, data monitoring, and outreach, with advanced visualization techniques in future HEP experiments.

% \bibliographystyle{nst}
% \bibliography{literature}
\printbibheading[heading=bibintoc]
% \printbibheading
\printbibliography[heading=none]
% \printbibliography
\end{document}